\documentclass[twocolumn,aps,prb, floatfix]{revtex4}
\usepackage{graphicx}
\usepackage{amsmath,verbatim}
\usepackage{tikz}
\usepackage{tcolorbox}
\usepackage{xcolor}
\usepackage{courier}
\begin{document}

\preprint{APS/123-QED}
\title{Phase-Field Modeling of Two-Phase Flows: A Projection-Based Cahn–Hilliard–Navier–Stokes Framework}

\author{Sukriti Manna}
\affiliation{Department of Mechanical and Industrial Engineering, University of Illinois, Chicago, IL 60607, USA \\
Center for Nanoscale Materials, Argonne National Laboratory, Lemont, IL 60439, USA \\
\texttt{smanna@uic.edu}}

\author{Constantine M Megaridis}
\affiliation{Department of Mechanical and Industrial Engineering, University of Illinois, Chicago, IL 60607, USA \\
\texttt{cmm@uic.edu}}

\author{Subramanian KRS Sankaranarayanan}
\affiliation{Department of Mechanical and Industrial Engineering, University of Illinois, Chicago, IL 60607, USA \\
Center for Nanoscale Materials, Argonne National Laboratory, Lemont, IL 60439, USA  \\
\texttt{skrssank@uic.edu}}

\begin{abstract}
The coupled Cahn--Hilliard and Navier--Stokes (CH--NS) equations provide a robust framework for modeling multiphase flows with diffuse interfaces, enabling the simulation of interfacial phenomena such as droplet breakup, bubble dynamics, and hydrodynamic instabilities. These capabilities are critical for applications including boiling heat transfer, microfluidics, coating processes, additive manufacturing, and oil--water separation, where accurate resolution of fluid--fluid interactions is essential. However, solving the CH--NS system is numerically challenging: the Cahn--Hilliard equation itself involves higher-order derivatives and nonlinear terms, and when coupled with the Navier--Stokes equations, strong two-way interactions emerge between interface motion and flow dynamics. In particular, the velocity field advects the phase-field variable, while the evolving interface modifies density and viscosity distributions that feed back into the flow, creating a tightly coupled nonlinear system. These interactions become even more complex in variable-density and variable-viscosity systems, where property contrasts across the interface must be resolved without compromising stability or mass conservation. To address these challenges, we employ a decoupled pressure projection method for solving the governing equations, with spatial discretization on staggered grids using a finite-difference scheme and explicit Euler integration in time. Our formulation extends the classical CH--NS system to both homogeneous and variable-property fluids, with careful treatment of the coupling between hydrodynamics and phase-field evolution. Validation against canonical benchmarks---including bubble rise and the Plateau--Taylor instability---shows excellent agreement in rise velocity, interface deformation, and instability wavelength. This framework provides a reproducible foundation for future extensions to multiphysics problems such as heat transfer, phase change, and electrohydrodynamic effects in applications spanning boiling, droplet manipulation, and thermal management in electronics.
\end{abstract}

\maketitle
\tableofcontents

\section{\label{sec:Intro}Introduction}
{
\setlength{\parskip}{0pt}
\setlength{\parindent}{1em}

Multiphase flows involving two or more immiscible fluids are central to a wide range of natural processes and engineering applications, including boiling heat transfer~\cite{li2025experimental}, microfluidics~\cite{raynaldo2024microchannel}, inkjet printing~\cite{wang2025numerical}, coating flows~\cite{li2024numerical}, additive manufacturing~\cite{liu2025multiphase}, and oil--water separation~\cite{wu2001intelligent}. The dynamics of these systems are governed by the interplay between fluid motion, interfacial tension, and property variations across the interface. Accurate prediction of interfacial evolution is therefore critical for understanding and optimizing such processes.

A variety of numerical strategies have been developed to model multiphase flows. In \textit{sharp-interface} methods such as the level set (LS)~\cite{gibou2018review} and volume of fluid (VOF)~\cite{gu2019volume}, the interface is assumed to have zero thickness. In contrast, \textit{diffuse-interface} (phase-field) methods represent the interface as a smooth transition region of finite thickness. This concept, dating back to Rayleigh~\cite{rayleigh1892} and van der Waals~\cite{van1979thermodynamic}, has since become a powerful and widely used tool for studying interfacial dynamics~\cite{steinbach2009phase,kim2012phase, lamorgese2011phase}. A key advantage of the diffuse-interface approach is that the governing equations can be derived from an energy-based variational framework consistent with hydrodynamic principles. However, ensuring that numerical schemes respect the discrete energy law is critical: violations lead to artificial dissipation or dispersion near the interface and can generate large numerical errors~\cite{wang2011energy}. Thus, despite its success, the development of accurate, energy-stable, and general-purpose multiphase flow solvers remains an active research area.

In the diffuse-interface framework, the Cahn--Hilliard equation~\cite{cahn1961spinodal} governs the evolution of the order parameter, naturally capturing interface motion, deformation, and topological transitions. When coupled with the Navier--Stokes equations, it yields the Cahn--Hilliard--Navier--Stokes (CH--NS) system, a tightly coupled nonlinear model in which the velocity field advects the phase field, while the evolving interface modifies density and viscosity distributions that feed back into the flow. This bidirectional coupling introduces significant challenges, particularly in variable-density and variable-viscosity systems.

Over the years, a variety of numerical methods have been proposed to address these challenges. Convex splitting schemes~\cite{han2015second} are unconditionally stable and widely used for the nonlinear chemical potential, though they can be computationally expensive. Shen and Yang~\cite{shen2010energy} developed energy-stable schemes for both constant- and variable-density CH--NS flows, while Liu et al.~\cite{liu2022novel} proposed a fully decoupled and unconditionally stable scheme for the Allen--Cahn--Navier--Stokes system. The scalar auxiliary variable (SAV) approach~\cite{shen2018scalar} introduces an auxiliary variable to reformulate the free energy, simplifying the discretization, but at the cost of modifying the original energy structure. Similarly, invariant energy quadratization (IEQ)~\cite{yang2018efficient} offers second-order accuracy for complex nonlinearities, though its link to the original energy dissipation law holds only under certain conditions~\cite{zhang2021remark}. Chen and Zhao~\cite{chen2020novel} introduced linear, second-order approaches with modified leap-frog time marching. Adaptive methods, such as those of Chen and Shen~\cite{chen2016efficient}, refine both space and time dynamically to better capture localized interfacial dynamics while maintaining stability.

For the Navier--Stokes component, a wide family of splitting and projection methods has been developed. Guermond and Salgado~\cite{guermond2009splitting} proposed a penalty-based splitting method requiring only a constant-coefficient Poisson solve per time step, while Guermond and Minev~\cite{guermond2019high} introduced adaptive third-order time stepping for incompressible flows. Guermond et al.~\cite{guermond2006overview} summarized major classes of projection schemes, including non-incremental pressure correction, standard incremental pressure correction, velocity-correction, and consistent splitting methods. Fractional-step approaches have also been extended to moving contact line problems with variable density and viscosity~\cite{gao2014efficient}, where convex splitting of the Cahn--Hilliard equation ensures stability under mild assumptions.

The CH--NS framework has been employed to study a wide range of canonical multiphase flow problems. Shen and Yang~\cite{shen2010energy} examined the dynamics of single rising bubbles in heavier media under both constant- and variable-density formulations, while Shen et al.~\cite{shin2020navier} investigated bubble deformation in shear flows using high-order polynomial free energies. Yang and Kim~\cite{yang2020novel} analyzed bubble dynamics with variable mobility, highlighting the role of transport properties in interfacial evolution. In addition, adaptive phase-field solvers~\cite{chen2016efficient} and spectral-element formulations~\cite{xiao2022spectral} have been developed to improve computational efficiency and accuracy. Benchmark studies of rising bubbles have also been carried out across multiple numerical approaches~\cite{hysing2009quantitative}, including finite-element implementations~\cite{aland2012benchmark}, providing valuable reference data for validation and comparison.

In this work, we consolidate and present a self-contained derivation of the CH--NS system for both homogeneous and variable density and viscosity two-fluid systems, together with a comprehensive description of our discretization strategies. Our implementation employs a finite-difference framework with staggered grids for spatial discretization and explicit 
Euler integration in time. A decoupled pressure-projection method is used to stabilize the hydrodynamic coupling. Emphasis is placed on mass conservation, numerical stability, and computational efficiency. To validate the solver, we present results for two benchmark problems: \emph{(i)} the rising bubble, and \emph{(ii)} the Plateau--Rayleigh instability. Quantitative comparisons with reference data demonstrate excellent agreement in rise velocity, interface deformation, and instability wavelength. This reproducible framework provides a solid foundation for future extensions to multiphysics problems such as heat 
transfer, phase change, and electrohydrodynamic effects~\cite{szczukiewicz2014proposed,
rothman1994lattice,azizian2019electrohydrodynamic}.

The remainder of this article is organized as follows. Sections~\ref{sec:Method}--\ref{sec:gov_eq} introduce the governing equations of the CH--NS system 
for both constant- and variable-density/viscosity formulations, together with the relevant dimensionless parameters. Section~\ref{sec:disc-strategy} describes the spatial and temporal discretization strategies, including the treatment of coupling terms and property variations. Section~\ref{sec:Results} presents the benchmark problems, numerical setup, and corresponding results, with comparisons to reference data. Finally, Section~\ref{sec:conclusion} summarizes the main findings and outlines potential directions for future work.

\begin{figure*}[t]
    \centering
    \includegraphics[width=0.75\textwidth]{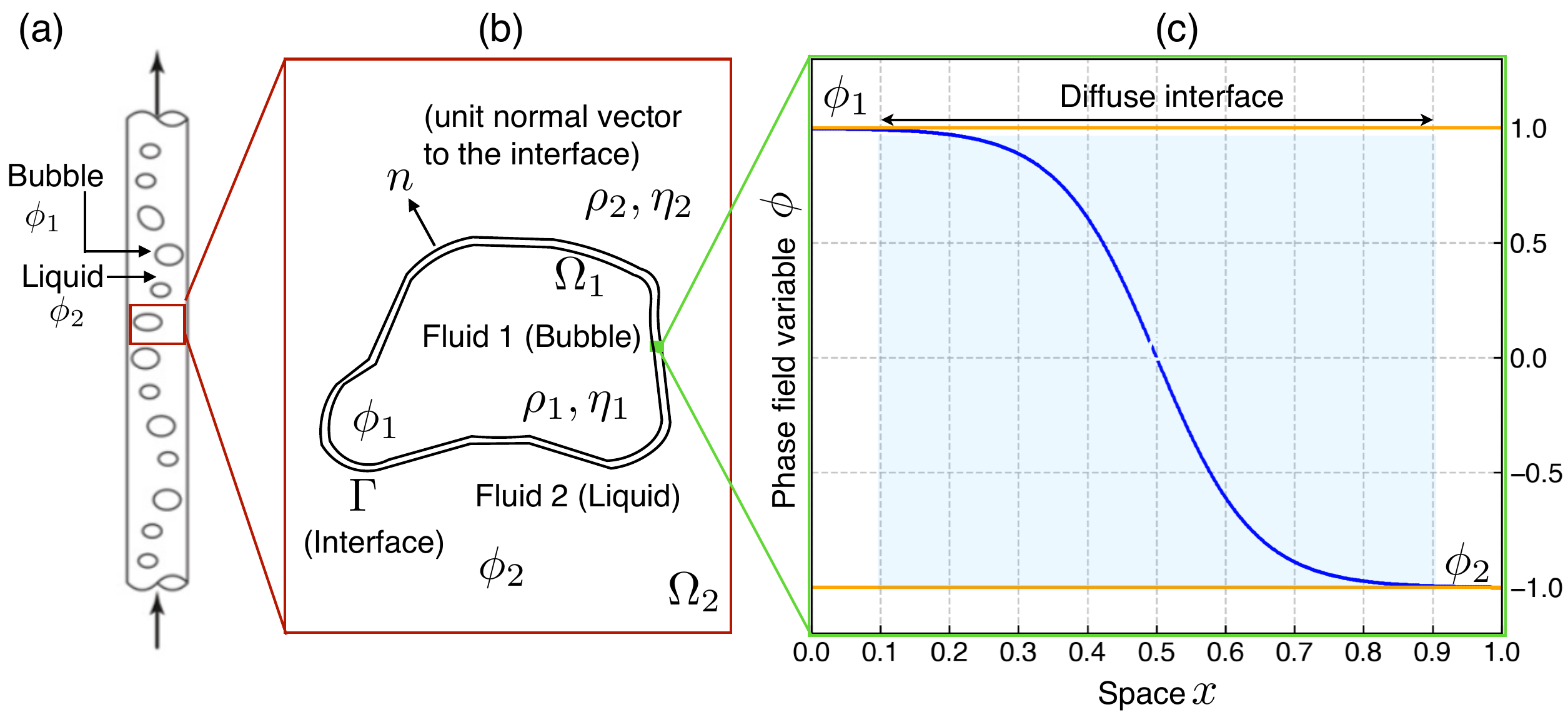}
  \caption{ Schematic representation of a diffuse interface between two fluids in a channel. (a) A flow of fluid containing a mixture of a gas bubble (Fluid 1) and a liquid (Fluid 2). (b) Amplified view of the bubble-liquid domain: $\Omega_1$ and $\Omega_2$ are the domains of Fluid 1 (Bubble) and Fluid 2 (Liquid), respectively; $\mathbf{n}$ is the unit normal vector to the interface $\Gamma$. (c) Phase field representation: $\phi$ is the phase field variable, with $\phi_1$ and $\phi_2$ representing values in Fluid 1 and Fluid 2, respectively; $x$ is the spatial coordinate. Properties across the interface: $\rho_1$ and $\eta_1$ are the density and viscosity of Fluid 1, while $\rho_2$ and $\eta_2$ are the density and viscosity of Fluid 2. The normal vector \(\mathbf{n}\) to the interface points from Fluid 2 towards Fluid 1, indicating the direction of the gradient of the phase field. The phase field variable \(\phi\) varies smoothly across the interface, transitioning from \(\phi_1\) in Fluid 1 to \(\phi_2\) in Fluid 2, with a finite thickness where \(\phi\) changes gradually, as shown in the blue shaded region. This continuous and differentiable nature of \(\phi\) allows for the modeling of complex interfacial phenomena.}
  \label{fig:schematics}
\end{figure*}

\section{\label{sec:Method}Methodology}
\subsection{\label{sec:PF_rep} Phase-Field Representation}
We consider a two-dimensional microchannel domain $\Omega$ containing two immiscible, incompressible, viscous fluids with constant densities $\rho_1$ and $\rho_2$ and dynamic viscosities $\eta_1$ and $\eta_2$, respectively. The spatial distribution of the two fluids is described by a phase-field variable $\phi(\mathbf{x},t)$, where $\mathbf{x}$ denotes the spatial coordinate and $t$ the time. The order parameter $\phi$ takes the bulk value $\phi=-1$ in Fluid~1 and $\phi=+1$ in Fluid~2, with a diffuse transition layer of finite thickness $\epsilon$ representing the fluid–fluid interface. 

Figure~\ref{fig:schematics} illustrates this diffuse-interface description. Panel (a) shows the microchannel geometry, where a gas bubble (Fluid~1) is surrounded by a liquid (Fluid~2). Panel (b) provides an enlarged view of the bubble–liquid system, highlighting the subdomains $\Omega_1$ and $\Omega_2$ occupied by Fluid~1 and Fluid~2, respectively. The interface $\Gamma$ separates the two fluids, with the unit normal vector $\mathbf{n}$ defined as pointing from the liquid (Fluid~2) toward the gas bubble (Fluid~1). Panel (c) depicts the smooth variation of the order parameter $\phi$ across the interfacial region (blue shaded layer), which enables the natural treatment of interfacial dynamics and material property variations without explicit interface tracking.

\subsection{\label{sec:PF_var}Definition of the Phase-Field Variable}
In the phase-field method, the interface emerges naturally from the continuous variation of an order parameter between two bulk values. For a binary incompressible mixture of a liquid (Fluid~1, also denoted A) and a gas (Fluid~2, also denoted B), each with distinct densities $\rho_1, \rho_2$ and viscosities $\eta_1, \eta_2$, the phase-field variable \cite{liu2003phase,kim2005continuous} can be defined in terms of the local masses $m_1$ and $m_2$ of the two fluids as
\begin{equation}
\phi = \frac{m_1 - m_2}{m_1 + m_2}, 
\qquad 
c = \frac{m_1}{m_1 + m_2},
\label{eq:phase_field}
\end{equation}
where $c$ denotes the volume fraction of Fluid~1 (liquid A). By construction, $\phi \in [-1,1]$ and $c \in [0,1]$. In this work, $\phi$ serves as the primary order parameter, while $c$ is used when convenient for interpretation.

\subsection{\label{sec:CHEq}Cahn--Hilliard Equation with Advection}
The temporal evolution of the phase field is governed by the Cahn--Hilliard (CH) equation~\cite{cahn1961spinodal}, which models the transport of a conserved scalar field subject to advection by the bulk velocity $\mathbf{u}$ and non-Fickian diffusion driven by gradients in chemical potential $\mu$:

\begin{equation}
\frac{\partial \phi}{\partial t} + \mathbf{u} \cdot \nabla \phi 
= \nabla \cdot \left[ -M(\phi) \nabla \mu \right],
\label{CH:advection}
\end{equation}

where $M(\phi) \ge 0$ is the mobility. 
The choice of $M(\phi)$---whether taken as a constant or made phase-dependent---determines whether bulk or interfacial diffusion dominates.

The chemical potential $\mu$ is obtained as the variational derivative of the total free energy functional (see Appendix~\ref{sec:free_energy} for the complete derivation).

\begin{equation}
\mathcal{E}(\phi) 
= \int_{\Omega} \left[ F(\phi) + \frac{\epsilon^2}{2} |\nabla \phi|^2 \right] \, d\mathbf{x},
\end{equation}
where $F(\phi) = \frac{1}{4}(\phi^2 - 1)^2$ is the double-well Helmholtz free energy density and $\epsilon$ is related to the interfacial thickness. The first term in $\mathcal{E}$ represents the bulk free energy, while the second term accounts for interfacial energy by penalizing sharp gradients in $\phi$.

For the concentration variable $c$, an equivalent form $F(c) = \frac14 c^2(c - 1)^2$ may be used. Both formulations capture phase separation and interfacial tension effects within the same energetic framework. The detailed derivation of Eq.~\eqref{CH:advection} are shown in Appendix~\ref{sec:CH-Adv Appendix}.

\begin{figure}[h]
    \centering
    \includegraphics[width=0.45\textwidth]{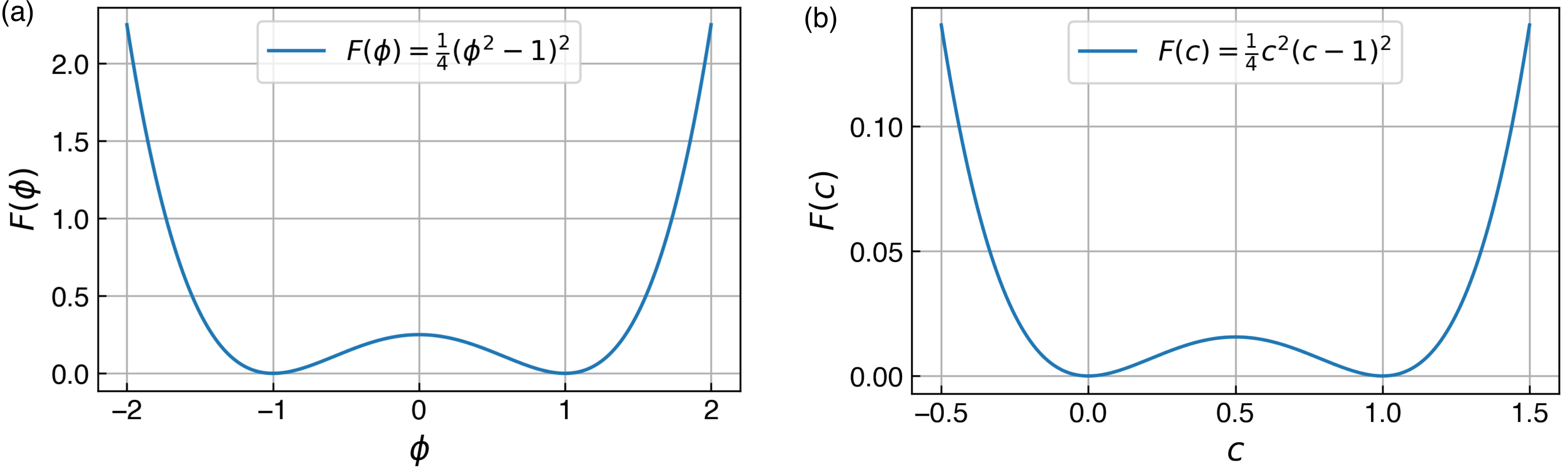}
    \caption{Free energy profiles for the phase field variable \(\phi\) and the concentration variable \(c\). (a) Free energy function \(F(\phi) = \frac{1}{4} (\phi^2 - 1)^2\) plotted against the phase field variable \(\phi\). This function represents the Helmholtz free energy of a unit volume of homogeneous material with composition \(\phi\). The graph illustrates how \(F(\phi)\) transitions smoothly, capturing the energy landscape between two immiscible phases characterized by \(\phi = -1\) and \(\phi = 1\). (b) Free energy function \(F(c) = \frac{1}{4} c^2 (c - 1)^2\) plotted against the concentration variable \(c\). This function represents the Helmholtz free energy related to the concentration variable in a binary mixture. The graph shows how \(F(c)\) changes with \(c\), illustrating the energy landscape between the phases characterized by \(c = 0\) and \(c = 1\). Both plots highlight the continuous and differentiable nature of the free energy functions, essential for modeling complex interfacial phenomena in binary fluid systems.}
    \label{fig:fre_energy_profile}
\end{figure}

\subsection{\label{sec:ChemPot}Chemical Potential in the Phase-Field Model}
Given the free-energy functional $\mathcal{E}(\phi)$ (Section~\ref{sec:free_energy}), the chemical potential $\mu$ follows from its variational derivative with respect to the order parameter $\phi$.

\begin{equation}
    \mu = \frac{\delta \mathcal{E}}{\delta \phi}.
    \label{eq:mu_def}
\end{equation}

The free energy functional is given by
\begin{equation}
    \mathcal{E}(\phi) = \int \left[ \frac{\epsilon^2}{2} |\nabla \phi|^2 
    + \frac{1}{4} (\phi^2 - 1)^2 \right] \, dV.
    \label{eq:free_energy}
\end{equation}

Let $\delta \phi$ be a small perturbation in $\phi$. The change in $\mathcal{E}(\phi)$ is
\begin{equation}
    \delta \mathcal{E} = \mathcal{E}(\phi + \delta \phi) - \mathcal{E}(\phi).
    \label{eq:delta_E_def}
\end{equation}

Substituting $\phi + \delta \phi$ into Eq.~\eqref{eq:free_energy}:
\begin{multline}
    \mathcal{E}(\phi + \delta \phi) 
    = \int \Bigg[ \frac{\epsilon^2}{2} 
    |\nabla (\phi + \delta \phi)|^2  \\
    + \frac{1}{4} \left( (\phi + \delta \phi)^2 - 1 \right)^2 \Bigg] \, dV.
    \label{eq:E_phi_plus_delta}
\end{multline}

For the gradient term:
\begin{align}
    |\nabla (\phi + \delta \phi)|^2 
    &= |\nabla \phi + \nabla \delta \phi|^2 \notag \\
    &= |\nabla \phi|^2 + 2 \nabla \phi \cdot \nabla \delta \phi 
    + |\nabla \delta \phi|^2.
    \label{eq:grad_expansion}
\end{align}
Neglecting the higher-order term $|\nabla \delta \phi|^2$:
\begin{equation}
    |\nabla (\phi + \delta \phi)|^2 \approx |\nabla \phi|^2 
    + 2 \nabla \phi \cdot \nabla \delta \phi.
    \label{eq:grad_approx}
\end{equation}

For the polynomial term:
\begin{multline}
    ((\phi + \delta \phi)^2 - 1)^2 
    = \left[ \phi^2 + 2 \phi \delta \phi + (\delta \phi)^2 - 1 \right]^2.
    \label{eq:poly_expansion}
\end{multline}
Neglecting $(\delta \phi)^2$:
\begin{equation}
    ((\phi + \delta \phi)^2 - 1)^2 
    \approx (\phi^2 - 1)^2 + 4 \phi (\phi^2 - 1) \, \delta \phi.
    \label{eq:poly_approx}
\end{equation}

Substituting these into $\mathcal{E}(\phi + \delta \phi)$:
\begin{multline}
    \mathcal{E}(\phi + \delta \phi) 
    \approx \int \Bigg[ \frac{\epsilon^2}{2} 
    \left( |\nabla \phi|^2 + 2 \nabla \phi \cdot \nabla \delta \phi \right) \\
    + \frac{1}{4} \left( (\phi^2 - 1)^2 + 4 \phi (\phi^2 - 1) \, \delta \phi \right) \Bigg] \, dV.
    \label{eq:E_phi_plus_delta_expanded}
\end{multline}

Subtracting $\mathcal{E}(\phi)$ from both sides:
\begin{equation}
    \delta \mathcal{E} \approx \int \left[ \epsilon^2 \nabla \phi \cdot \nabla \delta \phi 
    + \phi (\phi^2 - 1) \, \delta \phi \right] \, dV.
    \label{eq:delta_E_integral}
\end{equation}

Applying integration by parts to the gradient term and assuming $\delta \phi = 0$ on $\partial \Omega$:
\begin{equation}
    \int \epsilon^2 \nabla \phi \cdot \nabla \delta \phi \, dV 
    = -\int \epsilon^2 (\nabla^2 \phi) \, \delta \phi \, dV.
    \label{eq:int_by_parts}
\end{equation}

Here, $\nabla^2 \phi$ is the Laplacian:
\begin{equation}
    \nabla^2 \phi = \frac{\partial^2 \phi}{\partial x^2} 
    + \frac{\partial^2 \phi}{\partial y^2} 
    + \frac{\partial^2 \phi}{\partial z^2}.
    \label{eq:laplacian}
\end{equation}

Combining results:
\begin{equation}
    \delta \mathcal{E} = \int \left[ -\epsilon^2 \nabla^2 \phi 
    + \phi (\phi^2 - 1) \right] \delta \phi \, dV.
    \label{eq:delta_E_final}
\end{equation}

From the definition of the functional derivative:
\begin{equation}
    \frac{\delta \mathcal{E}}{\delta \phi} 
    = -\epsilon^2 \nabla^2 \phi + \phi (\phi^2 - 1).
    \label{eq:func_derivative}
\end{equation}

Thus, the chemical potential is:
\begin{equation}
    \mu = -\epsilon^2 \nabla^2 \phi + \phi (\phi^2 - 1).
    \label{eq:mu_final}
\end{equation}

\subsection{\label{sec:NV} Hydrodynamic Model: Navier--Stokes with Surface Tension}
Following the phase-field formulation and chemical potential introduced in Section~\ref{sec:ChemPot}, we couple the advective Cahn--Hilliard equation to the incompressible Navier--Stokes equations. Such coupling is a standard feature of diffuse-interface methods \cite{kim2012phase,anderson1998diffuse,he2008phase}, allowing the accurate treatment of interfacial dynamics 
in multiphase flows.

Let $\rho_1$ and $\rho_2$ denote the densities of the two fluids (bubble and liquid). In the phase-field framework, density and viscosity are expressed as linear functions\cite{liu2003phase} of the order parameter $\phi$:
\begin{align}
    \rho(\phi) &= \rho_1 \frac{1 + \phi}{2} + \rho_2 \frac{1 - \phi}{2}, 
    \label{eq:rho_phi} \\
    \eta(\phi) &= \eta_1 \frac{1 + \phi}{2} + \eta_2 \frac{1 - \phi}{2}.
    \label{eq:eta_phi}
\end{align}

With these definitions, the governing equations for an unsteady, viscous, incompressible, and immiscible two-fluid system take the form:
\begin{multline}
    \rho(\phi) \left[ \frac{\partial \mathbf{u}}{\partial t} 
    + (\mathbf{u} \cdot \nabla) \mathbf{u} \right] 
    = -\nabla p 
    + \nabla \cdot \left[ \eta(\phi) D(\mathbf{u}) \right] \\
    + \mathbf{F},
    \label{eq:NS_basic}
\end{multline}
where $\partial_t \mathbf{u}$ is the local acceleration, $(\mathbf{u} \cdot \nabla) \mathbf{u}$ is the convective acceleration, 
$-\nabla p$ is the pressure gradient force, $\nabla \cdot [\eta(\phi) D(\mathbf{u})]$ represents the viscous force per unit volume, 
and $\mathbf{F}$ accounts for external body forces (e.g., gravity). 

Here $D(\mathbf{u}) = \nabla \mathbf{u} + (\nabla \mathbf{u})^{T}$ is the rate-of-deformation tensor, i.e., 
the symmetric part of the velocity gradient.

At the interface $\Gamma$ between the two immiscible fluids (see Fig.~\ref{fig:schematics}), discontinuities in pressure and viscous stresses arise due to surface tension and property contrasts \cite{kim2012phase}. Surface tension balances the jump in normal stresses, ensuring mechanical equilibrium \cite{kim2012phase, uzgoren2009marker}. Incorporating this effect, the Navier--Stokes equations become:
\begin{multline}
    \rho(\phi) \left[ \frac{\partial \mathbf{u}}{\partial t} 
    + (\mathbf{u} \cdot \nabla) \mathbf{u} \right] 
    = -\nabla p 
    + \nabla \cdot \left[ \eta(\phi) D(\mathbf{u}) \right] \\
    + \mathbf{SF}_{\text{sing}} + \mathbf{F},
    \label{eq:NS_with_ST}
\end{multline}
where $\mathbf{SF}_{\text{sing}}$ denotes the singular surface tension force.

In our phase-field formulation, we adopt the surface tension model \cite{kim2005continuous}:
\begin{equation}
    \mathbf{SF}_{\text{sing}} = \tilde{\sigma} \, \varepsilon^{-1} \mu \nabla \phi,
    \label{eq:SF_sing}
\end{equation}
where $\tilde{\sigma}$ is a scaled surface tension parameter (proportional to the physical surface tension $\sigma$), $\mu$ is the chemical potential from Eq.~\eqref{eq:mu_final}, and $\varepsilon$ is the characteristic length scale over which the interface is diffused.

\subsection{\label{sec:gov_eq} Governing Equations in the Phase-Field Framework}
Extending the chemical potential formulation from the previous section, 
we couple the phase-field model to the Navier--Stokes equations\cite{jacqmin1999calculation} for two immiscible, viscous fluids. The resulting system enforces momentum conservation, 
incompressibility, and phase-field evolution, enabling a consistent description 
of bulk flow and interfacial physics.

\paragraph{Momentum conservation.}  

The momentum equation is derived from the incompressible Navier--Stokes equations. It accounts for density and viscosity variations through the phase field~$\phi$, 
and includes contributions from surface tension and external body forces:

\begin{align}
\rho(\phi) \left[ 
    \frac{\partial \mathbf{u}}{\partial t} 
    + (\mathbf{u} \cdot \nabla) \mathbf{u} 
\right] 
&= -\nabla p \notag \\
&\quad + \nabla \cdot \left[ \eta(\phi) \, D(\mathbf{u}) \right] \notag \\
&\quad + \mathbf{SF}_{\text{sing}} 
+ \mathbf{F},
\label{eq:NS_phasefield}
\end{align}

where $\rho(\phi)$ is the local density, $\mathbf{u}$ is the velocity field, $p$ is the pressure, 
$\eta(\phi)$ is the viscosity, $D(\mathbf{u})$ is the rate-of-deformation tensor, 
$\mathbf{SF}_{\text{sing}}$ is the singular surface tension force, 
and $\mathbf{F}$ represents external body forces such as gravity.

\paragraph{Incompressibility condition.}  
For incompressible flows, the velocity field must be divergence-free:
\begin{equation}
\nabla \cdot \mathbf{u} = 0,
\label{eq:incomp}
\end{equation}
ensuring conservation of mass in the fluid domain.

\paragraph{Phase-field evolution.}  
The interface motion is governed by the advective Cahn--Hilliard equation:
\begin{equation}
\frac{\partial \phi}{\partial t} + \mathbf{u} \cdot \nabla \phi
= \nabla \cdot \mathcal{J}, \label{eq:CH_adv}
\end{equation}
where $\mathcal{J}$ is the phase-field flux, given by
\begin{equation}
\mathcal{J} = -M(\phi) \nabla \mu. \label{eq:flux}
\end{equation}

\paragraph{Chemical potential.}  
Using the variational derivative of the free energy functional $\mathcal{E}(\phi)$ (see Section~\ref{sec:ChemPot}), the chemical potential is
\begin{equation}
\mu = -\epsilon^2 \nabla^2 \phi + \phi \, (\phi^2 - 1), \label{eq:mu_final2}
\end{equation}
where $\epsilon$ is the interface thickness parameter.

\paragraph{Surface tension force.}  
The singular surface tension force is modeled as
\begin{equation}
\mathbf{SF}_{\text{sing}} = \tilde{\sigma} \, \epsilon^{-1} \, \mu \, \nabla \phi, \label{eq:SF}
\end{equation}
where $\tilde{\sigma}$ is a scaled surface tension parameter proportional to the physical surface tension $\sigma$.

Equations~\eqref{eq:NS_phasefield}--\eqref{eq:SF} form the coupled Navier--Stokes--Cahn--Hilliard system used to simulate bubble dynamics in the present study.

\subsection{Property Variation Across the Interface}

As described in Section~\ref{sec:gov_eq}, the order parameter $\phi$ distinguishes the two phases:
\[
\phi = -1 \quad \text{(liquid phase)}, \qquad \phi = 1 \quad \text{(bubble phase)}.
\]
The densities $(\rho_1, \rho_2)$ and viscosities $(\eta_1, \eta_2)$ of these phases vary smoothly across the interface, and are obtained via linear interpolation (full derivation in Appendix~\ref{app:interp}).

\paragraph{Truncated phase field.}  
Solutions of the advective Cahn--Hilliard equation may yield $\phi$ values 
outside the physical range $[-1,1]$. This occurs because the equation does not strictly satisfy the maximum principle~\cite{shen2015decoupled}. 
To prevent unphysical overshoots, we define a truncated phase field:

\begin{equation}
\hat{\phi} =
\begin{cases}
\phi, & |\phi| \leq 1, \\
\mathrm{sign}(\phi), & |\phi| > 1,
\end{cases}
\label{eq:phi_trunc}
\end{equation}
where $\mathrm{sign}(\phi)$ returns $+1$ or $-1$.

\paragraph{Interpolated properties.}  
Replacing $\phi$ by $\hat{\phi}$ in the interpolation formulas gives:
\begin{align}
\eta(\phi) &= \frac{\hat{\phi} + 1}{2} \, (\eta_2 - \eta_1) + \eta_1,
\label{eq:eta_interp} \\
\rho(\phi) &= \frac{\hat{\phi} + 1}{2} \, (\rho_2 - \rho_1) + \rho_1.
\label{eq:rho_interp}
\end{align}

These expressions ensure smooth property variation across the diffuse interface while keeping $\phi$ within physical bounds.

\begin{figure}[h]
    \centering
    \includegraphics[width=0.45\textwidth]{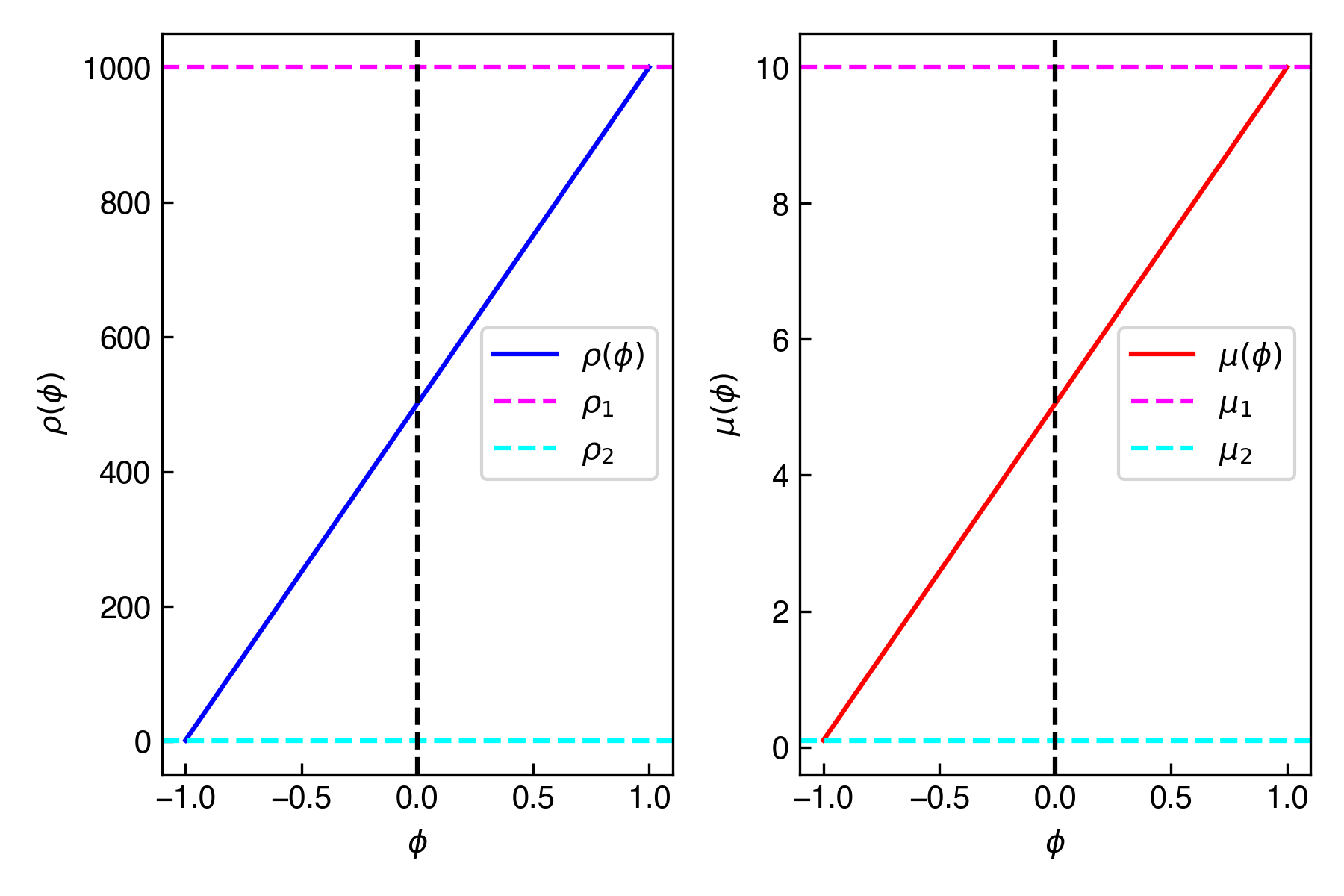}
    \caption{Variation of density $\rho(\phi)$ (left) and viscosity $\eta(\phi)$ (right) with the phase field parameter $\phi$. Both vary linearly between their respective phase values. The dashed line at $\phi = 0$ marks the interface.}
    \label{fig:mat_props}
\end{figure}

\subsection{\label{sec:disc-strategy} Discretization Strategy}
The coupled Cahn--Hilliard--Navier--Stokes (CH--NS) system governs both interfacial dynamics and bulk fluid motion in two-phase flow. Solving the fully coupled equations in a monolithic fashion is computationally demanding and can suffer from stability issues, particularly when large density or viscosity contrasts are present~\cite{guermond2006overview}. To improve numerical stability, reduce memory requirements, and efficiently leverage existing solvers, 
we employ a \emph{staggered time-stepping} (operator-splitting) scheme. 

At each time step $t^n \to t^{n+1}$, the solution procedure is divided into two stages:

1. \textbf{\emph{Momentum and pressure update.}}  
   The Navier--Stokes equations~\eqref{eq:NS_phasefield} are solved using the phase field 
   $\phi^n$ and chemical potential $\mu^n$ from the previous step, yielding the updated 
   velocity $\mathbf{u}^{n+1}$ and pressure $p^{n+1}$.

2. \textbf{\emph{Phase-field update.}}  
   With $\mathbf{u}^{n+1}$, the Cahn--Hilliard equation~\eqref{eq:CH_adv} is advanced to 
   compute $\phi^{n+1}$ and $\mu^{n+1}$, ensuring accurate interface transport and diffusion 
   while preserving incompressibility.

This fractional-step strategy decouples the CH and NS systems, allowing each subproblem to be treated with methods best suited to its characteristics. In practice, this approach maintains second-order temporal accuracy, good mass conservation, and robustness across a range of 
multiphase flow conditions~\cite{shen2010energy}.

\subsection{Time Discretization of the Momentum Equation}

In the first stage of the staggered update, we advance the velocity and pressure fields.  
The momentum conservation equation is

\begin{align}
\rho(\phi) \left( 
    \partial_t \mathbf{u} 
    + (\mathbf{u} \cdot \nabla) \mathbf{u} 
\right) 
&= -\nabla p \notag \\
&\quad + \nabla \cdot \left( \eta(\phi) D(\mathbf{u}) \right) \notag \\
&\quad + \mathbf{F} 
+ \tilde{\sigma} \varepsilon^{-1} \mu \nabla \phi,
\end{align}

where $\rho(\phi)$ and $\eta(\phi)$ are the phase-dependent density and viscosity, 
$\mathbf{u}$ is the velocity field, $p$ is the pressure, $D(\mathbf{u}) = \nabla \mathbf{u} + (\nabla \mathbf{u})^T$ 
is the rate-of-deformation tensor, $\mathbf{F}$ denotes external body forces (e.g., gravity), 
$\tilde{\sigma}$ is the scaled surface tension coefficient, $\varepsilon$ is the interface thickness parameter, 
$\mu$ is the chemical potential, and $\phi$ is the phase-field variable.  

We approximate the time derivative as
\begin{equation}
\partial_t \mathbf{u} \approx 
    \frac{\tilde{\mathbf{u}}^{n+1} - \mathbf{u}^n}{\Delta t},
\end{equation}
where $\mathbf{u}^n$ is the velocity at the current time step, 
$\tilde{\mathbf{u}}^{n+1}$ is an intermediate velocity field, 
and $\Delta t$ is the time step size.  

Substituting into the momentum equation gives
\begin{multline}
\rho(\phi) \left( 
    \frac{\tilde{\mathbf{u}}^{n+1} - \mathbf{u}^n}{\Delta t} 
    + (\mathbf{u}^n \cdot \nabla) \mathbf{u}^n 
\right) \\
= \underbrace{-\nabla p}_{\text{Term I}}
+ \underbrace{\nabla \cdot \left( \eta(\phi) D(\mathbf{u}^n) \right) 
+ \mathbf{F}^n 
+ \tilde{\sigma} \varepsilon^{-1} \mu^n \nabla \phi^n}_{\text{Term II}}.
\end{multline}

Here,  
\textbf{Term I}: pressure gradient contribution.  

\textbf{Term II}: viscous stress, external body forces, and surface tension effects.

We compute the intermediate velocity in four steps:

\noindent\textbf{Step 1: Intermediate velocity without pressure}  
Omitting \textbf{Term I}:
\begin{multline}
\rho(\phi) \left( \frac{\tilde{\mathbf{u}}^{n+1} - \mathbf{u}^n}{\Delta t} 
    + (\mathbf{u}^n \cdot \nabla) \mathbf{u}^n \right) = \\
    \nabla \cdot \left( \eta(\phi) D(\mathbf{u}^n) \right) 
    + \mathbf{F}^n 
    + \tilde{\sigma} \varepsilon^{-1} \mu^n \nabla \phi^n
\end{multline}
Rewriting:

\begin{align}
\rho(\phi) \frac{\tilde{\mathbf{u}}^{n+1} - \mathbf{u}^n}{\Delta t} 
&= \nabla \cdot \left( \eta(\phi) D(\mathbf{u}^n) \right) \notag \\
&\quad + \mathbf{F}^n 
+ \tilde{\sigma} \varepsilon^{-1} \mu^n \nabla \phi^n \notag \\
&\quad - \rho(\phi) (\mathbf{u}^n \cdot \nabla) \mathbf{u}^n
\end{align}

Multiplying by $\Delta t$:
\begin{align}
\rho(\phi) \left( \tilde{\mathbf{u}}^{n+1} - \mathbf{u}^n \right) 
&= \Delta t \Big[ \nabla \cdot \left( \eta(\phi) D(\mathbf{u}^n) \right) \notag \\
&\quad + \mathbf{F}^n 
+ \tilde{\sigma} \varepsilon^{-1} \mu^n \nabla \phi^n \notag \\
&\quad - \rho(\phi) (\mathbf{u}^n \cdot \nabla) \mathbf{u}^n \Big]
\end{align}

Thus:
\begin{multline}
\tilde{\mathbf{u}}^{n+1} 
    = \mathbf{u}^n 
    - \Delta t \left[ (\mathbf{u}^n \cdot \nabla) \mathbf{u}^n \right] \\
    + \frac{\Delta t}{\rho(\phi)} \left\{ \nabla \cdot \left[ \eta(\phi) D(\mathbf{u}^n) \right] 
    + \mathbf{F}^n 
    + \tilde{\sigma} \varepsilon^{-1} \mu^n \nabla \phi^n \right\}
\end{multline}
This intermediate $\tilde{\mathbf{u}}^{n+1}$ is not divergence-free.

\noindent\textbf{Step 2: Pressure correction}  
\begin{equation}
\rho(\phi) \frac{\mathbf{u}^{n+1} - \tilde{\mathbf{u}}^{n+1}}{\Delta t} 
    = -\nabla p^{n+1}
\end{equation}
Rearranging:
\begin{equation}
\mathbf{u}^{n+1} = \tilde{\mathbf{u}}^{n+1} 
    - \frac{\Delta t}{\rho(\phi)} \nabla p^{n+1}
\end{equation}

\noindent\textbf{Step 3: Enforcing incompressibility}  
Taking divergence and setting $\nabla \cdot \mathbf{u}^{n+1} = 0$:
\begin{equation}
\nabla \cdot \left( \frac{1}{\rho(\phi)} \nabla p^{n+1} \right) 
    = \frac{1}{\Delta t} \nabla \cdot \tilde{\mathbf{u}}^{n+1}
\end{equation}
with Neumann boundary condition:
\begin{equation}
\frac{\partial p^{n+1}}{\partial n} \bigg|_{\partial \Omega} = 0
\end{equation}

\noindent\textbf{Step 4: Time-dependent pressure equation}  
We can write:
\begin{equation}
\partial_t p - \nabla \cdot \left( \frac{1}{\rho(\phi)} \nabla p \right) 
    = - \nabla \cdot \frac{\tilde{\mathbf{u}}^{n+1}}{\Delta t}
\end{equation}
Using explicit Euler with $\Delta \tilde{t} = \Delta t / K$:
\begin{multline}
p^{n+\frac{k+1}{K}} = p^{n+\frac{k}{K}} + \Delta \tilde{t} \Bigg[ 
    \nabla \cdot \left( \frac{1}{\rho(\phi)} \nabla p^{n+\frac{k}{K}} \right) \\
    - \nabla \cdot \frac{\tilde{\mathbf{u}}^{n+1}}{\Delta t} \Bigg],
    \quad k = 0, \dots, K-1
\end{multline}
The resulting divergence-free $\mathbf{u}^{n+1}$ is then used in the second stage to advance the Cahn–Hilliard equation.

\subsection{Time Discretization for the Cahn--Hilliard Equation}

In the second stage of the staggered scheme, we update the phase field $\phi$ using the velocity field $\mathbf{u}^{n+1}$ computed from the Navier--Stokes step.  
The chemical potential at time step $n$ is given by:
\begin{equation}
\mu^n = -\varepsilon^2 \Delta \phi^n + (\phi^n)^3 - \phi^n.
\end{equation}

\textbf{Step 1: Explicit Euler update of the CH equation}  
We discretize the advective Cahn--Hilliard equation:
\begin{equation}
\frac{\phi^{n+1} - \phi^n}{\Delta t} + \mathbf{u}^{n+1} \cdot \nabla \phi^n = m \, \Delta \mu^n.
\end{equation}

Multiplying through by $\Delta t$:
\begin{equation}
\phi^{n+1} - \phi^n + \Delta t \left( \mathbf{u}^{n+1} \cdot \nabla \phi^n \right) 
    = \Delta t \left( m \, \Delta \mu^n \right).
\end{equation}

\textbf{Step 2: Isolating $\phi^{n+1}$}  
Rearranging:
\begin{equation}
\phi^{n+1} = \phi^n - \Delta t \left( \mathbf{u}^{n+1} \cdot \nabla \phi^n \right) 
    + \Delta t \left( m \, \Delta \mu^n \right).
\end{equation}

Here:  
- The first term $\phi^n$ represents the phase field from the previous time step.  
- The second term accounts for advective transport by $\mathbf{u}^{n+1}$.  
- The third term describes diffusive relaxation driven by the chemical potential gradient.

\textbf{Step 3: Time-step restriction}  
Due to the explicit treatment of the fourth-order spatial derivative in the 
Cahn--Hilliard equation, the time step must satisfy~\cite{eyre1998unconditionally}:
\begin{equation}
\Delta t < \frac{C \, h^4}{m \, \varepsilon^2},
\end{equation}
where $\Delta t$ is the time-step size, $h$ is the spatial grid spacing, $m$ is the mobility coefficient, $\varepsilon$ is the diffuse-interface thickness parameter, and $C$ is a stability constant of order unity.  

In practice, this stability restriction is often less severe than it appears.  
The interface thickness $\varepsilon$ is very small: for realistic fluid--fluid interfaces, the physical thickness is on the order of nanometers (except near the critical point). Directly resolving such thin interfaces in numerical simulations becomes infeasible once the characteristic domain length exceeds the micrometer scale. To overcome this, many phase-field simulations employ a 
\emph{numerically broadened interface}~\cite{jacqmin1999calculation}, which 
retains the correct macroscopic dynamics while permitting feasible grid sizes 
and time steps.

\subsection{\label{sec:space-disc} Space Discretization}
The time-discrete scheme presented in the previous section must be further discretized in space to enable numerical implementation.  
We employ a finite difference method (FDM) on an equidistant rectangular grid with uniform spacing \(h\) in both \(x\)- and \(y\)-directions.  

Although the discussion below is restricted to two dimensions for clarity, the method extends naturally to three dimensions without loss of generality.

The computational domain is defined as:
\[
\Omega = [0, L_x] \times [0, L_y],
\]
which is subdivided into \(N_x \times N_y\) square cells:
\[
N_x = \frac{L_x}{h}, \quad N_y = \frac{L_y}{h}.
\]
This structured arrangement provides a simple, memory-efficient, and highly parallelizable framework for solving the coupled Cahn–Hilliard–Navier–Stokes system.

\subsubsection{Scalar Variables on Cell Centers}

Scalar variables — such as the phase field \(\phi\), chemical potential \(\mu\), and pressure \(p\) — are stored at the \textbf{centers} of the computational cells.  This collocation allows straightforward computation of Laplacians and other scalar operators.  
To handle boundary conditions consistently, we extend the domain with one layer of \emph{ghost points} outside each physical boundary.

The discrete representation of a scalar field \(\phi(x, y)\) is given by:
\begin{equation}
\begin{aligned}
\phi_{i,j} &= \phi \!\left( \left( i - \frac{1}{2} \right) h, 
                          \; \left( j - \frac{1}{2} \right) h \right), \\[4pt]
&\quad i = 0, \ldots, N_x + 1, \\
&\quad j = 0, \ldots, N_y + 1,
\end{aligned}
\end{equation}
where \(\phi\) denotes a generic scalar quantity and the index ranges include the ghost points.

\subsubsection{Velocity Variables on a Staggered Grid}
When solving incompressible flow problems, a direct collocation of pressure and velocity can lead to \emph{odd–even decoupling} and spurious checkerboard patterns in the pressure field.  To avoid these numerical artifacts, we adopt a \emph{staggered Cartesian grid}, also known as the \emph{Marker-and-Cell (MAC) grid} \cite{ferziger2019computational, harlow1965numerical}.  

In this arrangement:
\begin{itemize}
    \item The \(u_0\) (or \(u\)) velocity component — aligned with the \(x\)-direction — is stored at the \emph{midpoints of vertical cell faces}.
    \item The \(u_1\) (or \(v\)) velocity component — aligned with the \(y\)-direction — is stored at the \emph{midpoints of horizontal cell faces}.
\end{itemize}
This placement allows for natural, centered differences when computing the divergence of velocity or the gradient of pressure, leading to improved accuracy and stability.

Formally, the discrete velocity locations are:

\begin{align}
u_{0,i,j} &= u_0 \!\bigg( 
    \left( i - 1 \right) h, \;
    \left( j - \tfrac{1}{2} \right) h 
\bigg), \nonumber\\[-4pt]
i &= 0, \ldots, N_x + 2, \quad 
j = 0, \ldots, N_y + 1, \\[6pt]
u_{1,i,j} &= u_1 \!\bigg( 
    \left( i - \tfrac{1}{2} \right) h, \;
    \left( j - 1 \right) h 
\bigg), \nonumber\\[-4pt]
i &= 0, \ldots, N_x + 1, \quad 
j = 0, \ldots, N_y + 2.
\end{align}

The index ranges include ghost points for the velocity variables, which simplify the treatment of velocity boundary conditions. 
Ghost points are artificial nodes introduced just outside the physical domain that allow uniform indexing of interior and boundary nodes. 
They enable the straightforward imposition of Dirichlet, Neumann, or periodic boundary conditions without special-case handling, 
and they eliminate conditional logic in stencil operations. The additional memory required is negligible compared with the benefits 
in implementation efficiency and clarity. 

In this work, we also employ a staggered, or Marker-and-Cell (MAC) as shown in Figure~\ref{fig:staggered_grid}, grid arrangement~\cite{harlow1965numerical}, 
where scalar quantities (\(\phi, \mu, p\)) are stored at cell centers, while velocity components are defined on cell faces 
(\(u_0\) at vertical faces and \(u_1\) at horizontal faces). This configuration offers several advantages: it improves stability by suppressing spurious pressure oscillations, ensures a more accurate coupling between divergence and gradient operators for enhanced mass conservation, and yields better-conditioned linear systems for pressure correction~\cite{guermond2006overview}. These properties are particularly important in the coupled Cahn--Hilliard--Navier--Stokes framework, where strong interaction between velocity, pressure, and interface dynamics is essential for both physical fidelity and numerical robustness. This spatial discretization framework forms the foundation for our numerical method:  scalar variables at cell centers ensure accurate phase field and pressure calculations,  while staggered velocities at cell faces enable stable and physically consistent fluid–structure interaction.

\begin{figure*}[t]
    \centering
    \includegraphics[width=0.75\textwidth]{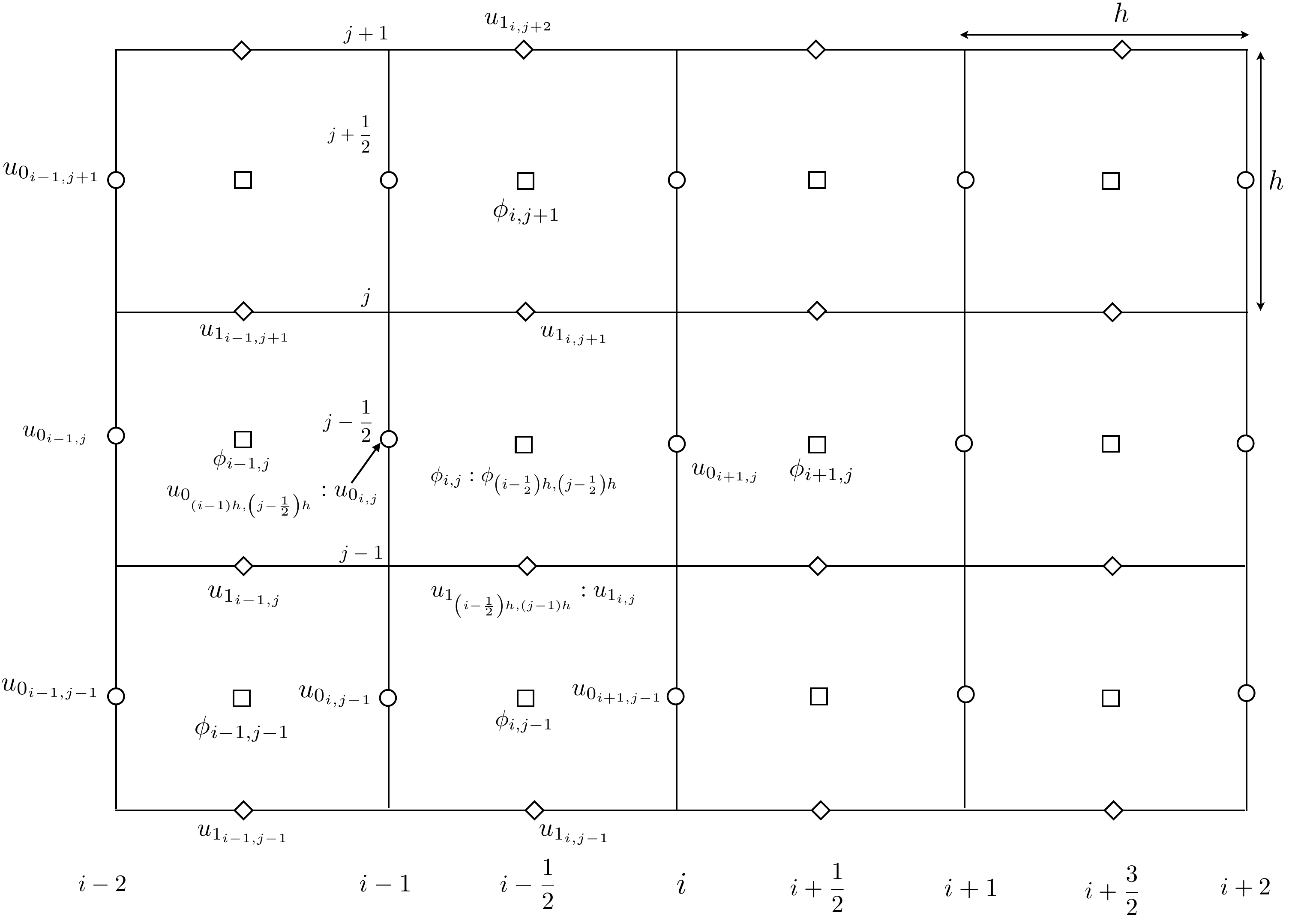}
    \caption{
        Illustration of the staggered (MAC) grid layout.  
        Scalar variables (\(\phi_{i,j}\), \(\mu_{i,j}\), \(p_{i,j}\)) are stored at cell centers.  
        The \(u_0\) velocity components are stored at the midpoints of vertical faces,  
        and the \(u_1\) velocity components are stored at the midpoints of horizontal faces.  
        Ghost points are included for both scalar and velocity variables, allowing unified indexing and consistent application of boundary conditions.  
        This arrangement enhances numerical stability and prevents spurious pressure–velocity decoupling.
    }
    \label{fig:staggered_grid}
\end{figure*}

In the following section, we describe the finite difference stencils used to approximate spatial derivatives on this grid.

\subsection{Constant Density and Viscosity}
\subsubsection{Space Discretization of the Momentum Equation}

We begin with the governing equation for the intermediate velocity field:
\begin{multline}
\tilde{\mathbf{u}}^{n+1} 
= \mathbf{u}^n 
- \Delta t \left[(\mathbf{u}^n \cdot \nabla) \mathbf{u}^n\right] \\
+ \frac{\Delta t}{\rho(\phi)} \left\{ \nabla \cdot \left[ \eta(\phi) D(\mathbf{u}^n) \right] 
+ \mathbf{F}^n 
+ \tilde{\sigma} \epsilon^{-1} \mu^n \nabla \phi^n \right\}.
\end{multline}

Under the assumption of constant density and viscosity, 
\(\eta(\phi) = \eta\) and \(\rho(\phi) = \rho\), 
the equation becomes:
\begin{multline}
\tilde{\mathbf{u}}^{n+1} 
= \mathbf{u}^n 
- \Delta t \underbrace{\left[(\mathbf{u}^n \cdot \nabla) \mathbf{u}^n \right]}_{\text{Term (i)}} \\
+ \frac{\Delta t}{\rho} \left\{ 
\underbrace{\nabla \cdot \left[ \eta D(\mathbf{u}^n) \right] + \mathbf{F}^n}_{\text{Term (ii)}} 
+ \underbrace{\tilde{\sigma} \epsilon^{-1} \mu^n \nabla \phi^n}_{\text{Term (iii)}} 
\right\}.
\end{multline}

Here, \(\mathbf{u} = [u_0, u_1]\) is the velocity vector in 2D.  
We next simplify Term (ii).

\paragraph{Rate-of-deformation tensor.}
The rate-of-deformation tensor \(D(\mathbf{u})\) is defined as:
\begin{equation}
D(\mathbf{u}) = \frac{1}{2} \left( \nabla \mathbf{u} + (\nabla \mathbf{u})^T \right).
\end{equation}
In Cartesian coordinates:
\begin{equation}
D(\mathbf{u}) 
= \frac{1}{2} 
\begin{pmatrix}
2 \frac{\partial u_0}{\partial x} & \frac{\partial u_0}{\partial y} + \frac{\partial u_1}{\partial x} \\
\frac{\partial u_1}{\partial x} + \frac{\partial u_0}{\partial y} & 2 \frac{\partial u_1}{\partial y}
\end{pmatrix}.
\end{equation}

The components may also be written as:
\begin{equation}
D(\mathbf{u}) = 
\begin{pmatrix}
\frac{\partial u_0}{\partial x} & \frac{1}{2} \left( \frac{\partial u_0}{\partial y} + \frac{\partial u_1}{\partial x} \right) \\
\frac{1}{2} \left( \frac{\partial u_1}{\partial x} + \frac{\partial u_0}{\partial y} \right) & \frac{\partial u_1}{\partial y}
\end{pmatrix}.
\end{equation}

\paragraph{Divergence of $D(\mathbf{u})$.}
Taking the divergence gives:
\begin{align}
\nabla \cdot D(\mathbf{u}) 
&= \frac{1}{2} 
\begin{pmatrix}
\frac{\partial^2 u_0}{\partial x^2} 
+ \tfrac{1}{4} \frac{\partial^2 u_0}{\partial y^2} 
+ \tfrac{1}{4} \frac{\partial^2 u_1}{\partial x \partial y} \\[6pt]
\tfrac{1}{2} \frac{\partial^2 u_0}{\partial x \partial y} 
+ \tfrac{1}{2} \frac{\partial^2 u_1}{\partial x^2} 
+ \frac{\partial^2 u_1}{\partial y^2}
\end{pmatrix} \\
&=
\begin{pmatrix}
\tfrac{1}{2}\frac{\partial^2 u_0}{\partial x^2} 
+ \tfrac{1}{8}\frac{\partial^2 u_0}{\partial y^2} 
+ \tfrac{1}{8}\frac{\partial^2 u_1}{\partial x \partial y} \\[6pt]
\tfrac{1}{4}\frac{\partial^2 u_0}{\partial x \partial y} 
+ \tfrac{1}{4}\frac{\partial^2 u_1}{\partial x^2} 
+ \tfrac{1}{2}\frac{\partial^2 u_1}{\partial y^2}
\end{pmatrix}.
\end{align}

Equivalently:
\begin{equation}
\nabla \cdot D(\mathbf{u}) = \frac{1}{2} 
\begin{pmatrix}
\frac{\partial^2 u_0}{\partial x^2} 
+ \frac{1}{4} \left( \frac{\partial^2 u_0}{\partial y^2} + \frac{\partial^2 u_1}{\partial x \partial y} \right) \\
\frac{1}{4} \left( \frac{\partial^2 u_0}{\partial x \partial y} + \frac{\partial^2 u_1}{\partial x^2} \right) 
+ \frac{\partial^2 u_1}{\partial y^2}
\end{pmatrix}.
\end{equation}

For incompressible flow, \(\nabla \cdot \mathbf{u} = 0\).
\begin{equation}
\frac{\partial u_0}{\partial x} + \frac{\partial u_1}{\partial y} = 0.
\end{equation}

This implies
\begin{align}
\frac{\partial^2 u_1}{\partial x \, \partial y} &= -\,\frac{\partial^2 u_0}{\partial y^2}, \label{eq:rel1}\\
\frac{\partial^2 u_0}{\partial x \, \partial y} &= -\,\frac{\partial^2 u_1}{\partial x^2}. \label{eq:rel2}
\end{align}

Substituting, the $x$-component becomes:
\begin{equation}
(\nabla \cdot D(\mathbf{u}))_x = \frac{1}{2} \frac{\partial^2 u_0}{\partial x^2},
\end{equation}
and the $y$-component:
\begin{equation}
(\nabla \cdot D(\mathbf{u}))_y = \frac{1}{2} \frac{\partial^2 u_1}{\partial y^2}.
\end{equation}

Thus:
\begin{equation}
\nabla \cdot D(\mathbf{u}) 
= \frac{1}{2} 
\begin{pmatrix}
\frac{\partial^2 u_0}{\partial x^2} \\
\frac{\partial^2 u_1}{\partial y^2}
\end{pmatrix}.
\end{equation}

\paragraph{Relation to the Laplacian.}  
For an incompressible velocity field $\mathbf{u} = (u_0,u_1)^T$, the vector Laplacian is
\begin{equation}
\nabla^2 \mathbf{u} =
\begin{pmatrix}
\dfrac{\partial^2 u_0}{\partial x^2} + \dfrac{\partial^2 u_0}{\partial y^2} \\[6pt]
\dfrac{\partial^2 u_1}{\partial x^2} + \dfrac{\partial^2 u_1}{\partial y^2}
\end{pmatrix}.
\end{equation}
Using the incompressibility condition, it follows that
\begin{equation}
\nabla \cdot D(\mathbf{u}) = \nabla^2 \mathbf{u}.
\end{equation}
Therefore, when the viscosity is constant,
\begin{equation}
\nabla \cdot \bigl(\eta D(\mathbf{u}^n)\bigr) = \eta \, \nabla^2 \mathbf{u}^n.
\end{equation}

\paragraph{Discrete Laplacian for $u_0$.}
\begin{equation}
\Delta_h u_{0_{i,j}} = \frac{u_{0_{i+1,j}} + u_{0_{i-1,j}} + u_{0_{i,j+1}} + u_{0_{i,j-1}} - 4u_{0_{i,j}}}{h^2}.
\end{equation}

\paragraph{Discrete Laplacian for $u_1$.}
\begin{equation}
\Delta_h u_{1_{i,j}} = \frac{u_{1_{i+1,j}} + u_{1_{i-1,j}} + u_{1_{i,j+1}} + u_{1_{i,j-1}} - 4u_{1_{i,j}}}{h^2}.
\end{equation}

\paragraph{Term (iii) -- Chemical potential term.}
In $x$-direction:
\begin{equation}
\frac{\tilde{\sigma}}{\varepsilon} \mu_{i,j}^n \nabla_x \phi_{i,j}^n 
\approx \frac{\tilde{\sigma}}{\varepsilon} \frac{\mu_{i,j}^n + \mu_{i-1,j}^n}{2} \cdot \frac{\phi_{i,j}^n - \phi_{i-1,j}^n}{h}.
\end{equation}
In $y$-direction:
\begin{equation}
\frac{\tilde{\sigma}}{\varepsilon} \mu_{i,j}^n \nabla_y \phi_{i,j}^n 
\approx \frac{\tilde{\sigma}}{\varepsilon} \frac{\mu_{i,j}^n + \mu_{i,j-1}^n}{2} \cdot \frac{\phi_{i,j}^n - \phi_{i,j-1}^n}{h}.
\end{equation}

\paragraph{Term (i) -- Convective term.}
The continuous form:
\begin{equation}
(\mathbf{u}^n \cdot \nabla) \mathbf{u}^n =
\begin{pmatrix}
u_0^n \frac{\partial u_0^n}{\partial x} + u_1^n \frac{\partial u_0^n}{\partial y} \\
u_0^n \frac{\partial u_1^n}{\partial x} + u_1^n \frac{\partial u_1^n}{\partial y}
\end{pmatrix}.
\end{equation}

\begin{widetext}
\begin{equation}
(\mathbf{u}^n \cdot \nabla \mathbf{u}^n)_{i,j} =
\begin{pmatrix}
\displaystyle
u_{0_{i,j}}^n \, 
\frac{u_{0_{i+1,j}}^n - u_{0_{i-1,j}}^n}{2h} 
+ \left( 
\frac{u_{1_{i,j+1}}^n + u_{1_{i-1,j+1}}^n 
      + u_{1_{i,j}}^n + u_{1_{i-1,j}}^n}{4} 
\right) 
\frac{u_{0_{i,j+1}}^n - u_{0_{i,j-1}}^n}{2h} \\[10pt]
\displaystyle
\left( 
\frac{u_{0_{i+1,j}}^n + u_{0_{i,j}}^n 
      + u_{0_{i+1,j-1}}^n + u_{0_{i,j-1}}^n}{4} 
\right) 
\frac{u_{1_{i+1,j}}^n - u_{1_{i-1,j}}^n}{2h} 
+ u_{1_{i,j}}^n \, 
\frac{u_{1_{i,j+1}}^n - u_{1_{i,j-1}}^n}{2h}
\end{pmatrix}.
\end{equation}
\end{widetext}

\paragraph{Final assembly}  
For each component \(\alpha \in \{0,1\}\), the update is:
\begin{widetext}
\begin{equation}
u_{\alpha,i,j}^{n+1} 
= u_{\alpha,i,j}^n
- \Delta t \,
\big[ (\mathbf{u}^n \cdot \nabla) u_\alpha^n \big]_{i,j}
+ \frac{\Delta t}{\rho} 
\Big[ \eta \, \Delta_h u_{\alpha,i,j}^n
+ f_{\alpha,i,j}^n
+ (\text{surface tension})_{\alpha,i,j} \Big].
\end{equation}
\end{widetext}

\subsubsection{Space Discretization of the Pressure Computation Equation}

We begin with the semi-implicit pressure update equation:
\begin{equation}
p^{n+\frac{k+1}{K}} 
= p^{n+\frac{k}{K}} 
+ \Delta \tilde{t} \left[ \nabla \cdot \left( \frac{1}{\rho(\phi)} \nabla p^{n+\frac{k}{K}} \right) 
- \nabla \cdot \frac{\tilde{\mathbf{u}}^{n+1}}{\Delta t} \right],
\end{equation}
where \(k=0,\dots,K-1\) denotes the sub-iterations within a single time step.

For a homogeneous system, \(\rho(\phi) = \rho\), \(\Delta t = dt\), and 
\(\Delta \tilde{t} = \frac{\rho h^2}{8}\), which yields:
\begin{equation}
p^{n+\frac{k+1}{K}} 
= p^{n+\frac{k}{K}} 
+ \frac{\rho h^2}{8} 
\left[ \nabla \cdot \left( \frac{1}{\rho} \nabla p^{n+\frac{k}{K}} \right) 
- \nabla \cdot \frac{\tilde{\mathbf{u}}^{n+1}}{dt} \right].
\end{equation}

Since \(\rho\) is constant, the diffusion term simplifies as:
\begin{equation}
\nabla \cdot \left( \frac{1}{\rho} \nabla p^{n+\frac{k}{K}} \right) 
= \frac{1}{\rho} \nabla^2 p^{n+\frac{k}{K}}.
\end{equation}

Substituting, we obtain:
\begin{equation}
p^{n+\frac{k+1}{K}} 
= p^{n+\frac{k}{K}} 
+ \frac{h^2}{8} \left[ \nabla^2 p^{n+\frac{k}{K}} 
- \rho \nabla \cdot \frac{\tilde{\mathbf{u}}^{n+1}}{dt} \right].
\end{equation}

\paragraph{Spatial discretization.}  
The Laplacian and divergence operators are expanded in two dimensions as:
\begin{align}
\nabla^2 p^{n+\frac{k}{K}} 
&= \frac{\partial^2 p^{n+\frac{k}{K}}}{\partial x^2} 
  + \frac{\partial^2 p^{n+\frac{k}{K}}}{\partial y^2}, \\
\nabla \cdot \tilde{\mathbf{u}}^{n+1} 
&= \frac{\partial \tilde{u}_x^{n+1}}{\partial x} 
  + \frac{\partial \tilde{u}_y^{n+1}}{\partial y}.
\end{align}

Including the \(\rho\) and \(dt\) scaling for the divergence term:
\begin{equation}
\rho \nabla \cdot \frac{\tilde{\mathbf{u}}^{n+1}}{dt} 
= \rho \left( \frac{\partial \tilde{u}_x^{n+1}}{\partial x} 
+ \frac{\partial \tilde{u}_y^{n+1}}{\partial y} \right) \frac{1}{dt}.
\end{equation}

\paragraph{Finite difference form.}  
The second-order central difference approximation for the Laplacian is:
\begin{align}
\nabla^2 p^{\,n+\tfrac{k}{K}} \;\approx\;& 
\frac{p_{i+1,j}^{\,n+\tfrac{k}{K}} - 2p_{i,j}^{\,n+\tfrac{k}{K}} + p_{i-1,j}^{\,n+\tfrac{k}{K}}}{h^2} \notag \\
&+ \frac{p_{i,j+1}^{\,n+\tfrac{k}{K}} - 2p_{i,j}^{\,n+\tfrac{k}{K}} + p_{i,j-1}^{\,n+\tfrac{k}{K}}}{h^2}.
\end{align}

The forward-difference approximation for the divergence is:
\begin{equation}
\nabla \cdot \tilde{\mathbf{u}}^{n+1} \approx 
\frac{\tilde{u}_{0,i+1,j}^{n+1} - \tilde{u}_{0,i,j}^{n+1}}{h} 
+ \frac{\tilde{u}_{1,i,j+1}^{n+1} - \tilde{u}_{1,i,j}^{n+1}}{h}.
\end{equation}

Hence:
\begin{equation}
\rho \nabla \cdot \frac{\tilde{\mathbf{u}}^{n+1}}{dt} \approx 
\rho \left( 
\frac{\tilde{u}_{0,i+1,j}^{n+1} - \tilde{u}_{0,i,j}^{n+1}}{h} 
+ \frac{\tilde{u}_{1,i,j+1}^{n+1} - \tilde{u}_{1,i,j}^{n+1}}{h} 
\right) \frac{1}{dt}.
\end{equation}

\begin{widetext}
\paragraph{Fully discretized pressure update.}  
Combining all terms, we have:
\begin{align}
p_{i,j}^{n+\frac{k+1}{K}} 
&= p_{i,j}^{n+\frac{k}{K}} 
+ \frac{h^2}{8} \left[ 
\frac{p_{i+1,j}^{n+\frac{k}{K}} - 2p_{i,j}^{n+\frac{k}{K}} + p_{i-1,j}^{n+\frac{k}{K}}}{h^2} 
+ \frac{p_{i,j+1}^{n+\frac{k}{K}} - 2p_{i,j}^{n+\frac{k}{K}} + p_{i,j-1}^{n+\frac{k}{K}}}{h^2} 
\right] \nonumber \\
&\quad - \frac{\rho h^2}{8} \left[ 
\frac{\tilde{u}_{0,i+1,j}^{n+1} - \tilde{u}_{0,i,j}^{n+1}}{h} 
+ \frac{\tilde{u}_{1,i,j+1}^{n+1} - \tilde{u}_{1,i,j}^{n+1}}{h} 
\right] \frac{1}{dt}.
\end{align}
\end{widetext}

\paragraph{Velocity correction step.}  
After the pressure update, the divergence-free velocity field is computed as:
\begin{equation}
\mathbf{u}^{n+1} = \tilde{\mathbf{u}}^{n+1} - \Delta t \frac{1}{\rho(\phi)} \nabla p^{n+1}.
\end{equation}

With:
\begin{equation}
\mathbf{u} = 
\begin{pmatrix} u_{0_{i,j}} \\ u_{1_{i,j}} \end{pmatrix},
\quad
\nabla p_{i,j}^{n+1} \approx 
\begin{pmatrix} 
\frac{p_{i,j}^{n+1} - p_{i-1,j}^{n+1}}{h} \\[4pt]
\frac{p_{i,j}^{n+1} - p_{i,j-1}^{n+1}}{h} 
\end{pmatrix}.
\end{equation}

The corrected velocity at each grid point is:
\begin{align}
u_{0_{i,j}}^{n+1} &= \tilde{u}_{0_{i,j}}^{n+1} - dt \frac{1}{\rho} \frac{p_{i,j}^{n+1} - p_{i-1,j}^{n+1}}{h}, \\
u_{1_{i,j}}^{n+1} &= \tilde{u}_{1_{i,j}}^{n+1} - dt \frac{1}{\rho} \frac{p_{i,j}^{n+1} - p_{i,j-1}^{n+1}}{h}.
\end{align}

In compact vector form:
\begin{equation}
\mathbf{u}_{i,j}^{n+1} = 
\begin{pmatrix} 
\tilde{u}_{0_{i,j}}^{n+1} - dt \frac{1}{\rho} \frac{p_{i,j}^{n+1} - p_{i-1,j}^{n+1}}{h} \\[4pt]
\tilde{u}_{1_{i,j}}^{n+1} - dt \frac{1}{\rho} \frac{p_{i,j}^{n+1} - p_{i,j-1}^{n+1}}{h} 
\end{pmatrix}.
\end{equation}

\subsubsection{Space Discretization of the Chemical Potential}  

The chemical potential is expressed as:
\begin{equation}
\mu = -\epsilon^2 \nabla^2 \phi + \phi (\phi^2 - 1),
\end{equation}
where \(\phi\) is the phase-field order parameter and \(\epsilon\) is a characteristic length scale.

\paragraph{Laplacian operator in two dimensions.}  
On a uniform grid with spacing \(h\) in both \(x\) and \(y\) directions, the Laplacian can be written as:
\begin{equation}
\nabla^2 \phi = \frac{\partial^2 \phi}{\partial x^2} + \frac{\partial^2 \phi}{\partial y^2}.
\end{equation}

Using the central difference approximation:
\begin{align}
\frac{\partial^2 \phi}{\partial x^2} &\approx \frac{\phi_{i+1,j} - 2\phi_{i,j} + \phi_{i-1,j}}{h^2}, \\
\frac{\partial^2 \phi}{\partial y^2} &\approx \frac{\phi_{i,j+1} - 2\phi_{i,j} + \phi_{i,j-1}}{h^2}.
\end{align}

Thus, the discrete Laplacian is:
\begin{align}
\nabla^2 \phi_{i,j} \;\approx\; &
\frac{\phi_{i+1,j} - 2\phi_{i,j} + \phi_{i-1,j}}{h^2} \notag \\
&+ \frac{\phi_{i,j+1} - 2\phi_{i,j} + \phi_{i,j-1}}{h^2}.
\end{align}

\paragraph{Discretized chemical potential.}  
Substituting the above discrete Laplacian into the expression for \(\mu\) gives:
\begin{align}
\mu_{i,j} &= 
-\epsilon^2 \Bigg[ 
\frac{\phi_{i+1,j} - 2\phi_{i,j} + \phi_{i-1,j}}{h^2} \notag \\
&\quad + \frac{\phi_{i,j+1} - 2\phi_{i,j} + \phi_{i,j-1}}{h^2} \Bigg] \notag \\
&\quad + \phi_{i,j} \left( \phi_{i,j}^2 - 1 \right).
\end{align}

This final expression represents the fully discretized chemical potential at grid point \((i,j)\), incorporating both the diffusive term (scaled by \(\epsilon^2\)) and the local nonlinear term \(\phi (\phi^2 - 1)\).

\subsubsection{Space Discretization of the Phase Field}

The evolution of the phase field (order parameter) \(\phi\) is governed by:
\begin{equation}
\phi_t + \mathbf{u} \cdot \nabla \phi = \nabla \cdot \mathcal{J}, \label{CH:adv1}
\end{equation}
where the flux \(\mathcal{J}\) is given by:
\begin{equation}
\mathcal{J} = M \nabla \mu,
\end{equation}
with \(M\) denoting the mobility and \(\mu\) the chemical potential.

\paragraph{Expanded form.}  
Substituting the flux expression into Eq.~\eqref{CH:adv} yields:
\begin{equation}
\phi_t + u_{0_{i,j}} \frac{\partial \phi}{\partial x} + u_{1_{i,j}} \frac{\partial \phi}{\partial y}
= \nabla \cdot (M \nabla \mu), \label{CH:adv_expanded}
\end{equation}
which, for constant \(M\), simplifies to:
\begin{equation}
\phi_t + u_{0_{i,j}} \frac{\partial \phi}{\partial x} + u_{1_{i,j}} \frac{\partial \phi}{\partial y}
= M \nabla^2 \mu.
\end{equation}

\paragraph{(a) Time derivative term.}  
Using a forward difference in time:
\begin{equation}
\phi_t \approx \frac{\phi_{i,j}^{n+1} - \phi_{i,j}^n}{\Delta t}.
\end{equation}

\paragraph{(b) Advection terms.}  

Using central differences in space:
\begin{align}
u_{0_{i,j}} \frac{\partial \phi}{\partial x} &\approx u_{0_{i,j}} \frac{\phi_{i+1,j} - \phi_{i-1,j}}{2h}, \\
u_{1_{i,j}} \frac{\partial \phi}{\partial y} &\approx u_{1_{i,j}} \frac{\phi_{i,j+1} - \phi_{i,j-1}}{2h}.
\end{align}

\paragraph{(c) Diffusion term.}  
\begin{align}
M \nabla^2 \mu \;\approx\;&\;
M \Bigg[
\frac{\mu_{i+1,j} - 2\mu_{i,j} + \mu_{i-1,j}}{h^2} \notag \\
&\quad + \frac{\mu_{i,j+1} - 2\mu_{i,j} + \mu_{i,j-1}}{h^2}
\Bigg].
\end{align}

\paragraph{Combined discretization.}  
\begin{widetext}
\begin{multline}
\frac{\phi_{i,j}^{n+1} - \phi_{i,j}^n}{\Delta t}
+ \left( \frac{u_{0_{i,j}} + u_{0_{i-1,j}}}{2} \right)
\frac{\phi_{i+1,j} - \phi_{i-1,j}}{2h}
+ \left( \frac{u_{1_{i,j}} + u_{1_{i,j-1}}}{2} \right)
\frac{\phi_{i,j+1} - \phi_{i,j-1}}{2h} \\
= M \left[
\frac{\mu_{i+1,j} - 2\mu_{i,j} + \mu_{i-1,j}}{h^2}
+ \frac{\mu_{i,j+1} - 2\mu_{i,j} + \mu_{i,j-1}}{h^2}
\right].
\end{multline}
\end{widetext}

\paragraph{Isolating \(\phi_{i,j}^{n+1}\).}  
\begin{widetext}
\begin{multline}
\phi_{i,j}^{n+1} - \phi_{i,j}^n
+ \Delta t \left[
\left( \frac{u_{0_{i,j}} + u_{0_{i-1,j}}}{2} \right)
\frac{\phi_{i+1,j} - \phi_{i-1,j}}{2h}
+ \left( \frac{u_{1_{i,j}} + u_{1_{i,j-1}}}{2} \right)
\frac{\phi_{i,j+1} - \phi_{i,j-1}}{2h}
\right] \\
= \Delta t M \left[
\frac{\mu_{i+1,j} - 2\mu_{i,j} + \mu_{i-1,j}}{h^2}
+ \frac{\mu_{i,j+1} - 2\mu_{i,j} + \mu_{i,j-1}}{h^2}
\right].
\end{multline}
\end{widetext}

\paragraph{Compact final form.}  
\begin{widetext}
\begin{multline}
\phi_{i,j}^{n+1} = \phi_{i,j}^n
- \frac{\Delta t}{2h} \left[
\left( \frac{u_{0_{i,j}} + u_{0_{i-1,j}}}{2} \right)
(\phi_{i+1,j} - \phi_{i-1,j})
+ \left( \frac{u_{1_{i,j}} + u_{1_{i,j-1}}}{2} \right)
(\phi_{i,j+1} - \phi_{i,j-1})
\right] \\
+ \frac{\Delta t M}{h^2} \left[
(\mu_{i+1,j} - 2\mu_{i,j} + \mu_{i-1,j})
+ (\mu_{i,j+1} - 2\mu_{i,j} + \mu_{i,j-1})
\right].
\end{multline}
\end{widetext}

\subsection{Spatial Discretization: Non-Matched Properties}

In the general case of non-matched density and viscosity, the discretization of the velocity update equation becomes more elaborate.  
We define:
\[
\eta_{i,j}^n = \eta(\phi_{i,j}^n), \quad
\rho_{i,j}^n = \rho(\phi_{i,j}^n).
\]

Starting from the velocity predictor:
\begin{widetext}
\begin{equation}
\tilde{\mathbf{u}}^{n+1} = \mathbf{u}^n 
- \Delta t \underbrace{\big[(\mathbf{u}^n \cdot \nabla) \mathbf{u}^n \big]}_{\text{Term (i)}}
+ \frac{\Delta t}{\rho} \left\{
\underbrace{\nabla \cdot \left[ \eta D(\mathbf{u}^n) \right] + \mathbf{F}^n}_{\text{Term (ii)}}
+ \underbrace{\tilde{\sigma} \epsilon^{-1} \mu^n \nabla \phi^n}_{\text{Term (iii)}}
\right\}.
\end{equation}
\end{widetext}

\paragraph{Viscous term for non-homogeneous viscosity.}
The rate-of-deformation tensor is:
\[
D(\mathbf{u}) = \frac{1}{2} \left( \nabla \mathbf{u} + (\nabla \mathbf{u})^T \right).
\]
In Cartesian coordinates:
\[
D(\mathbf{u}) =
\frac{1}{2}
\begin{pmatrix}
2 \frac{\partial u_0}{\partial x} & \frac{\partial u_0}{\partial y} + \frac{\partial u_1}{\partial x} \\
\frac{\partial u_1}{\partial x} + \frac{\partial u_0}{\partial y} & 2 \frac{\partial u_1}{\partial y}
\end{pmatrix}.
\]

Multiplying by \(2\eta\), we obtain:
\[
2\eta D(\mathbf{u}) =
\begin{pmatrix}
2 \eta \frac{\partial u_0}{\partial x} &
\eta \left( \frac{\partial u_0}{\partial y} + \frac{\partial u_1}{\partial x} \right) \\[4pt]
\eta \left( \frac{\partial u_1}{\partial x} + \frac{\partial u_0}{\partial y} \right) &
2 \eta \frac{\partial u_1}{\partial y}
\end{pmatrix}.
\]

\paragraph{Divergence of the viscous stress.}
The \(x\)-component is:
\begin{widetext}
\begin{align}
\left[ \nabla \cdot ( 2\eta D(\mathbf{u}) ) \right]_x &=
\frac{\partial}{\partial x} \left( 2\eta_{ij} \frac{\partial u_0}{\partial x} \right)
+ \frac{\partial}{\partial y} \left( \eta_{ij} \frac{\partial u_0}{\partial y} \right)
+ \frac{\partial}{\partial y} \left( \eta_{ij} \frac{\partial u_1}{\partial x} \right) \\
&= 2\left( \frac{\partial \eta_{ij}}{\partial x} \frac{\partial u_0}{\partial x}
+ \eta_{ij} \frac{\partial^2 u_0}{\partial x^2} \right)
+ \left( \frac{\partial \eta_{ij}}{\partial y} \frac{\partial u_0}{\partial y}
+ \eta_{ij} \frac{\partial^2 u_0}{\partial y^2} \right) \nonumber \\
&\quad + \left( \frac{\partial \eta_{ij}}{\partial y} \frac{\partial u_1}{\partial x}
+ \eta_{ij} \frac{\partial^2 u_1}{\partial x \partial y} \right). \nonumber
\end{align}
\end{widetext}

The \(y\)-component is:
\begin{widetext}
\begin{align}
\left[ \nabla \cdot ( 2\eta D(\mathbf{u}) ) \right]_y &=
\frac{\partial}{\partial x} \left( \eta_{ij} \frac{\partial u_1}{\partial x} \right)
+ \frac{\partial}{\partial x} \left( \eta_{ij} \frac{\partial u_0}{\partial y} \right)
+ \frac{\partial}{\partial y} \left( 2\eta_{ij} \frac{\partial u_1}{\partial y} \right) \\
&= \left( \frac{\partial \eta_{ij}}{\partial x} \frac{\partial u_1}{\partial x}
+ \eta_{ij} \frac{\partial^2 u_1}{\partial x^2} \right)
+ \left( \frac{\partial \eta_{ij}}{\partial x} \frac{\partial u_0}{\partial y}
+ \eta_{ij} \frac{\partial^2 u_0}{\partial x \partial y} \right) \nonumber \\
&\quad + 2\left( \frac{\partial \eta_{ij}}{\partial y} \frac{\partial u_1}{\partial y}
+ \eta_{ij} \frac{\partial^2 u_1}{\partial y^2} \right). \nonumber
\end{align}
\end{widetext}

\paragraph{Full tensor form.}
\begin{widetext}
\[
\nabla \cdot \left( 2 \eta D(\mathbf{u}) \right) =
\begin{pmatrix}
2 \left( \frac{\partial \eta_{ij}}{\partial x} \frac{\partial u_0}{\partial x} + \eta_{ij} \frac{\partial^2 u_0}{\partial x^2} \right)
+ \left( \frac{\partial \eta_{ij}}{\partial y} \frac{\partial u_0}{\partial y} + \eta_{ij} \frac{\partial^2 u_0}{\partial y^2} \right)
+ \left( \frac{\partial \eta_{ij}}{\partial y} \frac{\partial u_1}{\partial x} + \eta_{ij} \frac{\partial^2 u_1}{\partial x \partial y} \right) \\[6pt]
\left( \frac{\partial \eta_{ij}}{\partial x} \frac{\partial u_1}{\partial x} + \eta_{ij} \frac{\partial^2 u_1}{\partial x^2} \right)
+ \left( \frac{\partial \eta_{ij}}{\partial x} \frac{\partial u_0}{\partial y} + \eta_{ij} \frac{\partial^2 u_0}{\partial x \partial y} \right)
+ 2 \left( \frac{\partial \eta_{ij}}{\partial y} \frac{\partial u_1}{\partial y} + \eta_{ij} \frac{\partial^2 u_1}{\partial y^2} \right)
\end{pmatrix}.
\]
\end{widetext}

\paragraph{Spatial discretization — $x$-component.}

\begin{widetext}
For the term \(2 ( \partial_x \eta_{ij} \, \partial_x u_0 + \eta_{ij} \, \partial^2_x u_0 )\):
\[
\partial_x \eta_{ij} \approx \frac{\eta_{i+1,j} - \eta_{i-1,j}}{2h}, \quad
\partial_x u_0 \approx \frac{u_{0,i+1,j} - u_{0,i-1,j}}{2h}, \quad
\eta_{ij} \, \partial^2_x u_0 \approx \eta_{ij} \frac{u_{0,i+1,j} - 2u_{0,i,j} + u_{0,i-1,j}}{h^2}.
\]
\end{widetext}
Analogous formulas hold for the \(y\)-derivatives and cross-derivatives.

The discretized \(x\)-component becomes:
\begin{widetext}
\begin{multline}
\left[ \nabla \cdot ( 2\eta D(\mathbf{u}) ) \right]_{x_{i,j}} =
2\left( \frac{\eta_{i+1,j} - \eta_{i-1,j}}{2h} \cdot \frac{u_{0,i+1,j} - u_{0,i-1,j}}{2h}
+ \eta_{ij} \frac{u_{0,i+1,j} - 2u_{0,i,j} + u_{0,i-1,j}}{h^2} \right) \\
+ \left( \frac{\eta_{i,j+1} - \eta_{i,j-1}}{2h} \cdot \frac{u_{0,i,j+1} - u_{0,i,j-1}}{2h}
+ \eta_{ij} \frac{u_{0,i,j+1} - 2u_{0,i,j} + u_{0,i,j-1}}{h^2} \right) \\
+ \left( \frac{\eta_{i,j+1} - \eta_{i,j-1}}{2h} \cdot \frac{u_{1,i+1,j} - u_{1,i-1,j}}{2h}
+ \eta_{ij} \frac{u_{1,i+1,j+1} - u_{1,i+1,j-1} - u_{1,i-1,j+1} + u_{1,i-1,j-1}}{4h^2} \right).
\end{multline}
\end{widetext}

\paragraph{Spatial discretization — $y$-component.}
Similarly:
\begin{widetext}
\begin{multline}
\left[ \nabla \cdot ( 2\eta D(\mathbf{u}) ) \right]_{y_{i,j}} =
\left( \frac{\eta_{i+1,j} - \eta_{i-1,j}}{2h} \cdot \frac{u_{1,i+1,j} - u_{1,i-1,j}}{2h}
+ \eta_{ij} \frac{u_{1,i+1,j} - 2u_{1,i,j} + u_{1,i-1,j}}{h^2} \right) \\
+ \left( \frac{\eta_{i+1,j} - \eta_{i-1,j}}{2h} \cdot \frac{u_{0,i,j+1} - u_{0,i,j-1}}{2h}
+ \eta_{ij} \frac{u_{0,i+1,j+1} - u_{0,i+1,j-1} - u_{0,i-1,j+1} + u_{0,i-1,j-1}}{4h^2} \right) \\
+ 2\left( \frac{\eta_{i,j+1} - \eta_{i,j-1}}{2h} \cdot \frac{u_{1,i,j+1} - u_{1,i,j-1}}{2h}
+ \eta_{ij} \frac{u_{1,i,j+1} - 2u_{1,i,j} + u_{1,i,j-1}}{h^2} \right).
\end{multline}
\end{widetext}

\subsection{Pressure Correction for Variable Density and Viscosity}

In the case of non-homogeneous density \(\rho(\mathbf{x},t)\) and viscosity, the pressure equation must be modified to account for spatial variations in these quantities. The iterative pressure update at sub-step \(k\) of the fractional-step method is written as:

\begin{widetext}
\begin{equation}
p^{n + \frac{k+1}{K}} = p^{n + \frac{k}{K}} 
+ \frac{\rho h^2}{8} \left[ 
\underbrace{\nabla \cdot \left( \frac{1}{\rho} \nabla p^{n + \frac{k}{K}} \right)}_{\text{Term (i)}} 
- \underbrace{\nabla \cdot \left( \frac{\tilde{\mathbf{u}}^{n+1}}{\Delta t} \right)}_{\text{Term (ii)}} 
\right]
\end{equation}
\end{widetext}

\subsubsection{Expansion of Term (i)}

Multiplying Term (i) by \(\rho^n\), we write:

\begin{equation}
\left( \rho^n \nabla \cdot \left( \frac{1}{\rho^n} \nabla p^{n+\frac{k}{K}} \right) \right)_{i,j}
\end{equation}

Expanding the divergence operator:

\begin{equation}
\rho^n \left[ 
\frac{\partial}{\partial x} \left( \frac{1}{\rho^n} \frac{\partial p^{n+\frac{k}{K}}}{\partial x} \right)
+ \frac{\partial}{\partial y} \left( \frac{1}{\rho^n} \frac{\partial p^{n+\frac{k}{K}}}{\partial y} \right)
\right]
\end{equation}

Using the product rule, we obtain:

\begin{align}
\left( \rho^n \nabla \cdot \left( \frac{1}{\rho^n} \nabla p^{n+\tfrac{k}{K}} \right) \right)_{i,j} 
&= \Big[ \Delta p^{n+\tfrac{k}{K}} \notag \\
&\quad + \rho^n \, \nabla \!\left( \frac{1}{\rho^n} \right) \cdot \nabla p^{n+\tfrac{k}{K}} \Big]_{i,j}.
\end{align}

\paragraph*{Discrete Laplacian \(\Delta p^{n+\frac{k}{K}}\):}

\begin{align}
\Delta p_{i,j}^{n+\tfrac{k}{K}} 
&= \frac{p_{i+1,j}^{\,n+\tfrac{k}{K}} - 2p_{i,j}^{\,n+\tfrac{k}{K}} + p_{i-1,j}^{\,n+\tfrac{k}{K}}}{h^2} \notag \\
&\quad + \frac{p_{i,j+1}^{\,n+\tfrac{k}{K}} - 2p_{i,j}^{\,n+\tfrac{k}{K}} + p_{i,j-1}^{\,n+\tfrac{k}{K}}}{h^2}.
\end{align}

\paragraph*{Gradient term \(\rho^n \nabla(1/\rho^n) \cdot \nabla p^{n+\frac{k}{K}}\):}

Central differences for \(\nabla (1/\rho^n)\):

\begin{equation}
\nabla \left( \frac{1}{\rho^n} \right)_{i,j} = 
\begin{pmatrix}
\frac{\frac{1}{\rho_{i+1,j}^n} - \frac{1}{\rho_{i-1,j}^n}}{2h} \\[6pt]
\frac{\frac{1}{\rho_{i,j+1}^n} - \frac{1}{\rho_{i,j-1}^n}}{2h}
\end{pmatrix}
\end{equation}

Central differences for \(\nabla p\):

\begin{equation}
\nabla p_{i,j}^{n+\frac{k}{K}} = 
\begin{pmatrix}
\frac{p_{i+1,j}^{n+\frac{k}{K}} - p_{i-1,j}^{n+\frac{k}{K}}}{2h} \\[6pt]
\frac{p_{i,j+1}^{n+\frac{k}{K}} - p_{i,j-1}^{n+\frac{k}{K}}}{2h}
\end{pmatrix}
\end{equation}

Combining these:

\begin{widetext}
\begin{equation}
\rho_{i,j}^n \nabla \left( \frac{1}{\rho_{i,j}^n} \right) \cdot \nabla p_{i,j}^{n+\frac{k}{K}} =
\rho_{i,j}^n \left[
\frac{\frac{1}{\rho_{i+1,j}^n} - \frac{1}{\rho_{i-1,j}^n}}{2h} 
\frac{p_{i+1,j}^{n+\frac{k}{K}} - p_{i-1,j}^{n+\frac{k}{K}}}{2h}
+ \frac{\frac{1}{\rho_{i,j+1}^n} - \frac{1}{\rho_{i,j-1}^n}}{2h} 
\frac{p_{i,j+1}^{n+\frac{k}{K}} - p_{i,j-1}^{n+\frac{k}{K}}}{2h}
\right]
\end{equation}
\end{widetext}

\paragraph*{Complete discrete form of Term (i):}

\begin{widetext}
\begin{equation}
\begin{aligned}
&\left[ \rho^n \nabla \cdot \left( \frac{1}{\rho^n} \nabla p^{n+\frac{k}{K}} \right) \right]_{i,j} =
\frac{p_{i+1,j}^{n+\frac{k}{K}} - 2p_{i,j}^{n+\frac{k}{K}} + p_{i-1,j}^{n+\frac{k}{K}}}{h^2}
+ \frac{p_{i,j+1}^{n+\frac{k}{K}} - 2p_{i,j}^{n+\frac{k}{K}} + p_{i,j-1}^{n+\frac{k}{K}}}{h^2} \\[4pt]
&\quad + \rho_{i,j}^n \left[
\frac{\frac{1}{\rho_{i+1,j}^n} - \frac{1}{\rho_{i-1,j}^n}}{2h}
\frac{p_{i+1,j}^{n+\frac{k}{K}} - p_{i-1,j}^{n+\frac{k}{K}}}{2h}
+ \frac{\frac{1}{\rho_{i,j+1}^n} - \frac{1}{\rho_{i,j-1}^n}}{2h}
\frac{p_{i,j+1}^{n+\frac{k}{K}} - p_{i,j-1}^{n+\frac{k}{K}}}{2h}
\right]
\end{aligned}
\end{equation}
\end{widetext}

\subsubsection{Expansion of Term (ii)}

We have:
\begin{equation}
\rho \nabla \cdot \frac{\tilde{\mathbf{u}}^{n+1}}{\Delta t} 
= \rho \left( \frac{\partial \tilde{u}_x^{n+1}}{\partial x} 
+ \frac{\partial \tilde{u}_y^{n+1}}{\partial y} \right) \frac{1}{\Delta t}
\end{equation}

Discrete form:
\begin{equation}
\nabla \cdot \tilde{\mathbf{u}}^{n+1} \approx
\frac{\tilde{u}_{0,i+1,j}^{n+1} - \tilde{u}_{0,i,j}^{n+1}}{h}
+ \frac{\tilde{u}_{1,i,j+1}^{n+1} - \tilde{u}_{1,i,j}^{n+1}}{h}
\end{equation}

Thus:
\begin{equation}
\rho \nabla \cdot \frac{\tilde{\mathbf{u}}^{n+1}}{\Delta t} \approx
\frac{\rho}{\Delta t} \left[
\frac{\tilde{u}_{0,i+1,j}^{n+1} - \tilde{u}_{0,i,j}^{n+1}}{h}
+ \frac{\tilde{u}_{1,i,j+1}^{n+1} - \tilde{u}_{1,i,j}^{n+1}}{h}
\right]
\end{equation}

The contribution to the pressure equation is then:
\begin{equation}
-\frac{\rho h^2}{8\Delta t} \left[
\frac{\tilde{u}_{0,i+1,j}^{n+1} - \tilde{u}_{0,i,j}^{n+1}}{h}
+ \frac{\tilde{u}_{1,i,j+1}^{n+1} - \tilde{u}_{1,i,j}^{n+1}}{h}
\right]
\end{equation}

\subsection{Inhomogeneous Density Correction in Divergence-Free Velocity Evaluation}

When density varies spatially, the velocity correction step becomes:
\begin{equation}
\mathbf{u}^{n+1} = \tilde{\mathbf{u}}^{n+1} - \Delta t \frac{1}{\rho(\phi)} \nabla p^{n+1}
\end{equation}

In discrete form:
\begin{equation}
\mathbf{u}_{i,j}^{n+1} =
\begin{pmatrix}
\tilde{u}_{0,i,j}^{n+1} - \frac{\Delta t}{\rho} \frac{p_{i,j}^{n+1} - p_{i-1,j}^{n+1}}{h} \\
\tilde{u}_{1,i,j}^{n+1} - \frac{\Delta t}{\rho} \frac{p_{i,j}^{n+1} - p_{i,j-1}^{n+1}}{h}
\end{pmatrix}
\end{equation}

Replacing \(\rho\) by local averages:
\begin{equation}
\mathbf{u}_{i,j}^{n+1} =
\begin{pmatrix}
\tilde{u}_{0,i,j}^{n+1} - \frac{\Delta t}{A_x \rho_{i,j}} \frac{p_{i,j}^{n+1} - p_{i-1,j}^{n+1}}{h} \\
\tilde{u}_{1,i,j}^{n+1} - \frac{\Delta t}{A_y \rho_{i,j}} \frac{p_{i,j}^{n+1} - p_{i,j-1}^{n+1}}{h}
\end{pmatrix}
\end{equation}

Where:
\begin{align}
A_x \rho_{i,j} &= \frac{\rho_{i,j} + \rho_{i-1,j}}{2} \\
A_y \rho_{i,j} &= \frac{\rho_{i,j} + \rho_{i,j-1}}{2}
\end{align}

Explicitly:
\begin{align}
u_{0,i,j}^{n+1} &= \tilde{u}_{0,i,j}^{n+1} - \frac{\Delta t}{\frac{\rho_{i,j} + \rho_{i-1,j}}{2}} \frac{p_{i,j}^{n+1} - p_{i-1,j}^{n+1}}{h} \\
u_{1,i,j}^{n+1} &= \tilde{u}_{1,i,j}^{n+1} - \frac{\Delta t}{\frac{\rho_{i,j} + \rho_{i,j-1}}{2}} \frac{p_{i,j}^{n+1} - p_{i,j-1}^{n+1}}{h}
\end{align}

\begin{figure*}[t]
    \centering
    \includegraphics[width=0.75\textwidth]{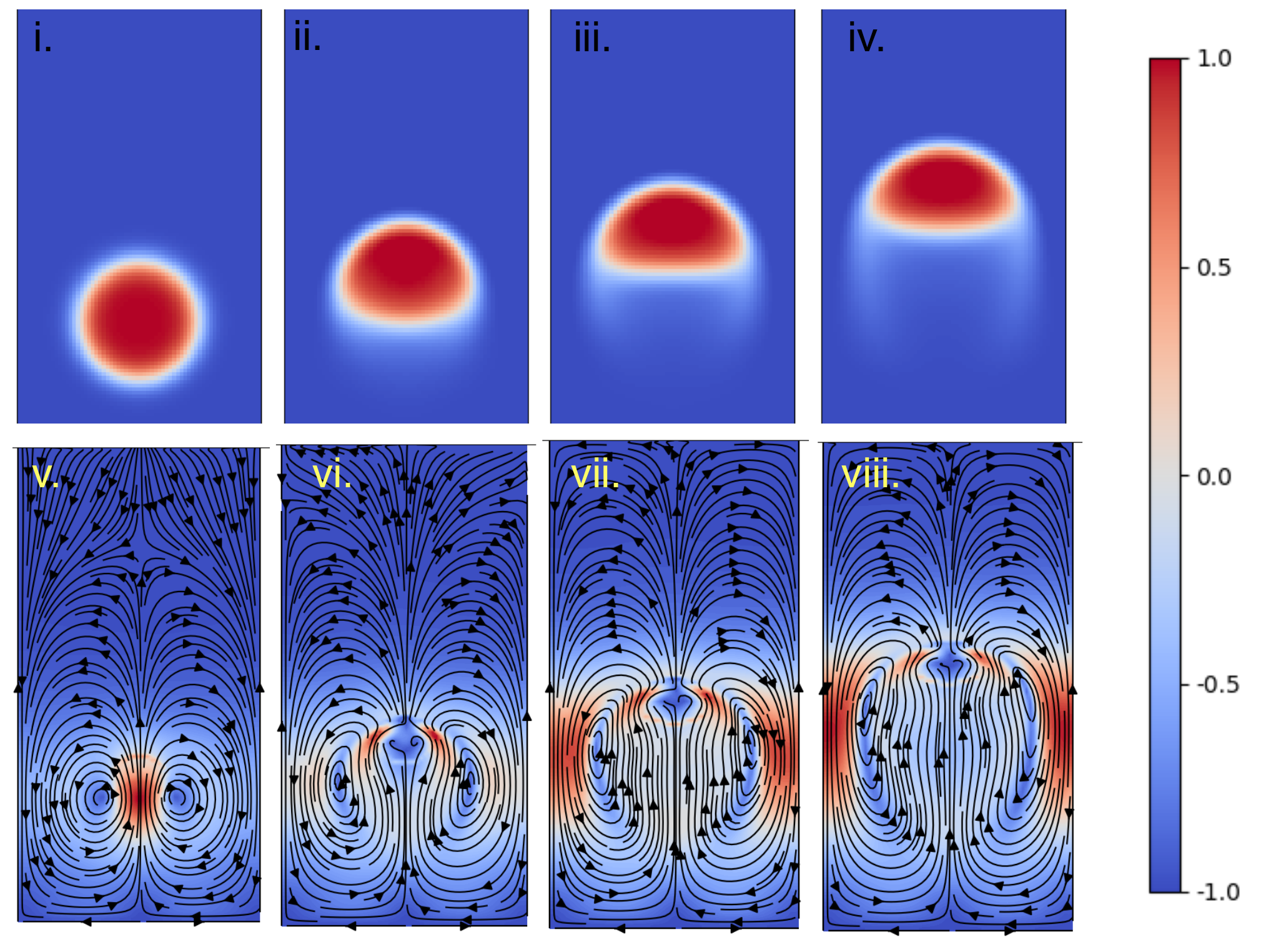}
    \caption{Rise of a single bubble in a confined microchannel using the phase-field method.  
    Top row (i–iv): Evolution of $\phi$ showing the bubble interface as it rises due to buoyancy.  
    Bottom row (v–viii): Velocity field and vorticity contours illustrating flow structures and counter-rotating vortices in the bubble wake.}
    \label{fig:bubble}
\end{figure*}

\begin{figure}[t]
    \centering
    \includegraphics[width=0.35\textwidth]{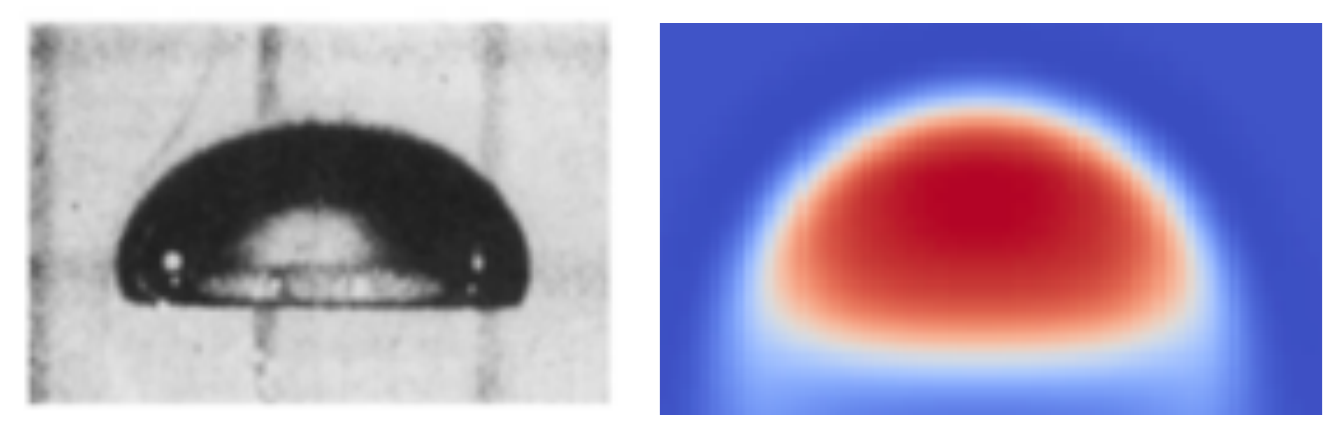}
    \caption{Comparison between experimental and simulated dome-shaped bubble profiles in a microchannel. The close agreement in interface curvature and morphology validates the numerical model. The left figure is reprinted with permission from Ref.~\cite{bhaga1981bubbles}.}
    \label{fig:bubble_comp_exp}
\end{figure}

\section{\label{sec:Results}Results \& Discussion}

Using the numerical framework outlined in Section~\ref{sec:Method}, we investigate two benchmark problems:  
(a) the rise of a single bubble in a non-homogeneous liquid, and  
(b) the Plateau–Rayleigh instability.  
These cases were selected because they test different aspects of the phase-field formulation: interfacial deformation under buoyancy in confined geometries, and capillary-driven instability in slender liquid structures.

\subsection{\label{sec:rising_bubble}Rising Bubble with Density and Viscosity Contrast}
The rise of bubbles in immiscible fluids is a canonical problem in multiphase flow, with relevance to microfluidic devices, chemical reactors, and natural systems such as gas release from sediments. In confined microchannels, bubble motion is governed by three competing effects:  
(a) \emph{Buoyancy}, which drives upward motion,  
(b) \emph{Surface tension}, which resists deformation and promotes minimal interface curvature, and  
(c) \emph{Viscous forces}, which damp motion and can induce asymmetric deformation near channel walls.  

In this work, we investigate these interactions using our coupled Cahn--Hilliard--Navier--Stokes solver, validated against benchmark rising-bubble tests, to capture the balance of buoyancy, surface tension, and viscous stresses in confined geometries.

\paragraph{Physical setup and initial conditions.}  
The computational domain is a rectangular region $\Omega = [1,0] \times [2,0]$ with uniform grid spacing $h$ in both $x$ and $y$ directions. The domain is initially filled with liquid ($\phi \approx -1$), and a bubble of radius $0.25$ is centered at $(0.5,\, 0.5)$ with $\phi \approx 1$.  
The initial phase-field distribution is given by:

\begin{align}
\phi_{i,j}^0 &= 
\tanh \Bigg(
    \frac{0.25 - \sqrt{ 
        \big((i - 0.5)h - 0.5\big)^2 } }
         {\sqrt{2\epsilon}} \nonumber \\
&\qquad\quad
    - \frac{ \big((j - 0.5)h - 0.5\big)^2 }
         {\sqrt{2\epsilon}}
\Bigg)
\end{align}

ensuring a smooth diffuse interface of width proportional to $\epsilon$.

Material properties as shown in Figure~\ref{fig:mat_props} are defined as:  
\[
\rho_{-1} = 1000, \quad \eta_{-1} = 10, \quad
\rho_{+1} = 100, \quad \eta_{+1} = 1,
\]
where the subscripts $-1$ and $+1$ refer to the outer liquid and the inner vapor phase, respectively.  
The gravitational force is specified as:
\[
\mathbf{f} = \big(0, -0.98 \,[\rho(\phi) - \rho_{-1}]\big),
\]
and the surface tension coefficient is $\sigma = 24.5$.  
Boundary conditions are \emph{no-slip} at the top and bottom walls and \emph{free-slip} at the lateral boundaries, mimicking a vertically confined microchannel.

\paragraph{Simulation results and flow features.}  
Figure~\ref{fig:bubble} (top row, i–iv) shows the evolution of the bubble interface as $\phi$ transitions smoothly between $\phi=1$ (vapor) and $\phi=-1$ (liquid). At $t=0$, the bubble is perfectly circular. As buoyancy accelerates the lighter vapor phase upward, deformation begins:  
- \emph{Step 250}: Slight elongation along the vertical axis, with the lower interface flattening due to upward acceleration.  
- \emph{Step 750}: The bubble exhibits a pronounced dome-like shape, with increased curvature near the top and compression at the bottom.  
- \emph{Step 1000}: The deformation becomes more substantial, reflecting the balance between buoyancy-driven stretching and surface tension’s restorative effect.

The bottom row (v–viii) in Figure~\ref{fig:bubble} presents velocity vectors overlaid on vorticity contours. Two symmetric counter-rotating vortices form in the bubble’s wake soon after motion begins. These vortices grow in size and intensity before stabilizing, generating a low-pressure zone behind the bubble.  
This vortex-induced pressure drop modifies the drag force, affecting the rise velocity and shape evolution. The gradual change in vortex strength over time illustrates the interplay between inertia and viscous dissipation in confined two-phase flows.

\paragraph{Validation with experimental data.}  
Figure~\ref{fig:bubble_comp_exp} compares the simulated bubble shape with experimental microchannel images~\cite{bhaga1981bubbles}. Both show a dome-shaped bubble with similar curvature, aspect ratio, and wall proximity effects. The agreement demonstrates that:  
1. The diffuse-interface formulation accurately captures \emph{interfacial deformation} and \emph{capillary effects} in confined geometries.  
2. The coupled hydrodynamic–phase-field approach reproduces \emph{wake-induced flow patterns} consistent with experimental visualization.  
3. The simulation maintains numerical stability and mass conservation over long timescales, essential for predictive multiphase modeling.

Overall, this test confirms the model’s robustness for simulating buoyancy-driven interfacial dynamics in confined microscale channels.

\begin{figure*}[t]
    \centering
    \includegraphics[width=0.8\textwidth]{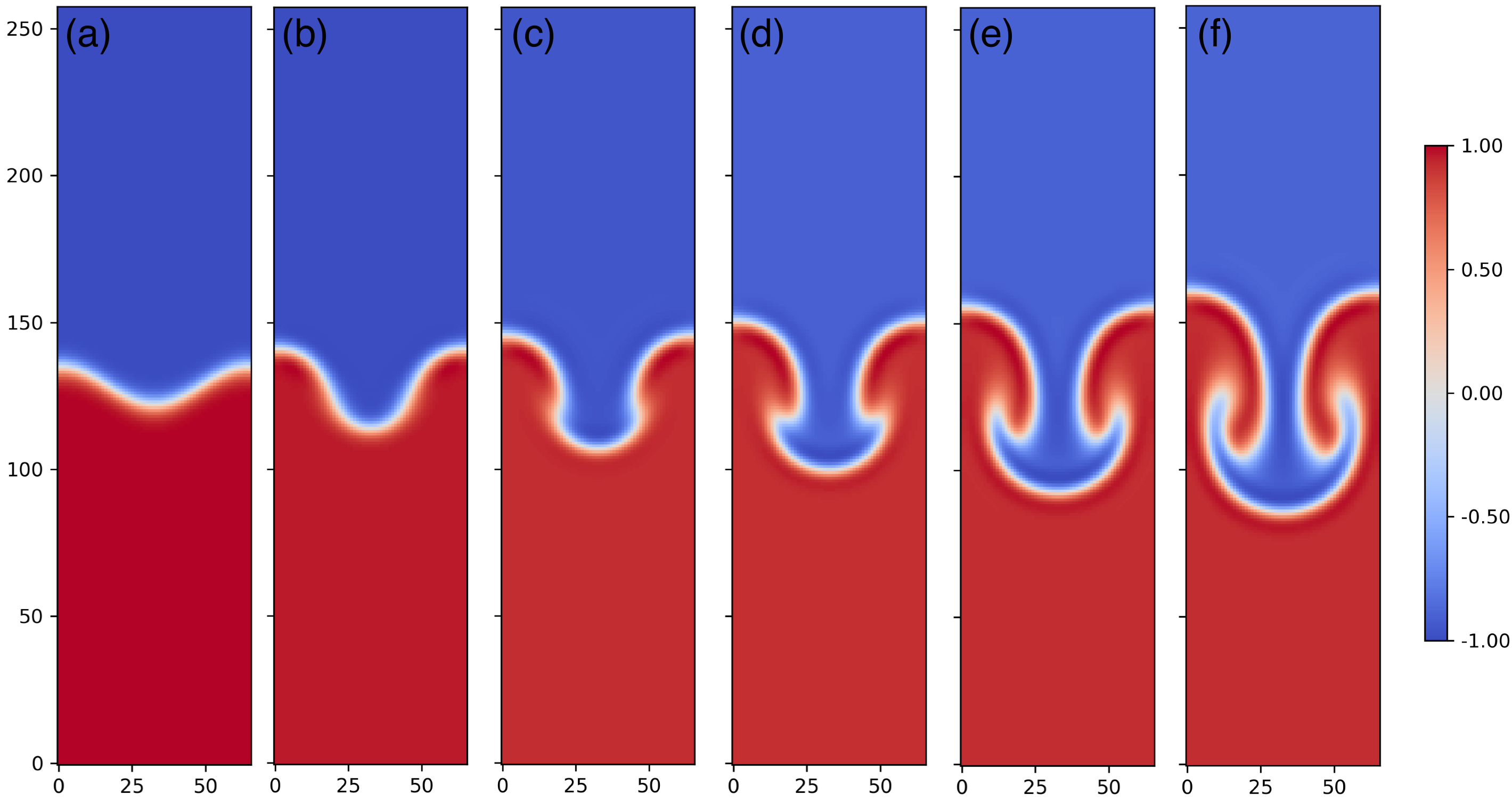}
    \caption{Time evolution of the Rayleigh–Taylor instability using the developed phase-field formulation. Snapshots (a)–(f) show the interface between a heavy fluid (red) overlying a lighter fluid (blue). Gravity amplifies the initial perturbation, producing downward-moving spikes of heavy fluid and upward-moving bubbles of light fluid, accompanied by vorticity generation and mixing.}
    \label{fig:PR}
\end{figure*}
\subsection{\label{sec:PRI}Rayleigh–Taylor Instability}

We now apply the phase-field formulation to the Rayleigh–Taylor instability (RTI), a canonical benchmark for studying buoyancy-driven interfacial instabilities in multiphase flows. RTI occurs when a denser fluid is placed above a lighter fluid in a gravitational field, making the configuration unstable. Small perturbations at the interface grow over time as the heavier fluid descends into the lighter one (spikes) and the lighter fluid rises into the heavier one (bubbles). This instability is relevant to a wide range of natural and engineering processes, including oceanic mixing, astrophysical flows (e.g., supernova remnants), and inertial confinement fusion.

\paragraph{Physical mechanism.}
The instability is driven by the gravitational potential energy difference between the two fluid layers. Any small perturbation to the initially flat interface is amplified because gravity accelerates the heavier fluid downward and the lighter fluid upward. Surface tension can act to stabilize short-wavelength disturbances, while viscosity modifies the growth rate and morphology.

\paragraph{Numerical setup.}
The computational domain is a rectangular box of size $\Omega = [0, L_x] \times [0, L_y]$, initially filled with two immiscible fluids. The initial phase-field distribution is given by:
\[
\phi(x,y,0) = \tanh\left( \frac{y - y_0 - 0.1\cos(2\pi x / L_x)}{\sqrt{2\epsilon}} \right),
\]
where $y_0$ denotes the initial interface height, and the sinusoidal perturbation of amplitude $0.1$ seeds the instability. The density and viscosity are defined as $\rho_{+1}$, $\eta_{+1}$ for the heavy fluid ($\phi \approx 1$) and $\rho_{-1}$, $\eta_{-1}$ for the light fluid ($\phi \approx -1$). The gravitational force is applied in the negative $y$-direction, $\mathbf{f} = (0, -g(\rho(\phi) - \rho_{-1}))$, with $g$ the gravitational acceleration. No-slip boundary conditions are imposed at the top and bottom, and free-slip boundaries at the sides. The coupled Navier–Stokes–Cahn–Hilliard system (Section~\ref{sec:Method}) governs the evolution.

\paragraph{Simulation results.}
Figure~\ref{fig:PR} presents the time evolution of the instability in six snapshots (a)–(f). Initially, the interface is nearly flat, with only the imposed sinusoidal perturbation visible. 
As time progresses, the heavier fluid penetrates downward in spike-like structures, while the lighter fluid rises upward in bubble-like protrusions. 
At early times, the perturbation amplitude grows exponentially, consistent with linear Rayleigh--Taylor theory~\cite{rayleigh1882investigation,taylor1950instability}. 
As the system enters the nonlinear regime, spikes sharpen and accelerate under gravity, while bubbles broaden as they rise~\cite{youngs1984numerical}. 
The shear that develops along the flanks of spikes and bubbles generates counter-rotating vortices, which in turn intensify and promote the formation of a mixing zone where small-scale interpenetration of the fluids occurs~\cite{youngs1984numerical}.

\paragraph{Discussion.}
The phase-field method captures both the early linear growth and the highly nonlinear deformation without requiring explicit interface tracking. The diffuse-interface representation ensures smooth curvature evaluation and accurate surface tension forces. The simulated patterns, including spike and bubble morphology, agree qualitatively with experimental and theoretical predictions of RTI evolution. This validates the capability of the present formulation to model buoyancy-driven instabilities with strong density and viscosity contrasts.

\section{Conclusion}
\label{sec:conclusion}
In this study, we employed the phase-field method within the framework of the coupled Navier–Stokes–Cahn–Hilliard (NSCH) system\cite{jacqmin1999calculation} to investigate the dynamics of immiscible two-phase flows. Two representative benchmark problems were considered: (i) the rise of a bubble in a non-homogeneous liquid, and (ii) the Rayleigh–Taylor instability between two fluids of different densities.

For the rising-bubble case, the simulations accurately reproduced the full sequence of dynamic behavior, from the initial spherical shape to progressive deformation under the competing influences of buoyancy, surface tension, and viscous drag. The formation and evolution of counter-rotating vortices in the wake, as well as the gradual shape transition to a dome-like interface, were captured with high fidelity, underscoring the method’s ability to resolve subtle interfacial curvature effects in confined geometries.

For the Rayleigh--Taylor instability, the model reproduced both the qualitative morphology and the quantitative growth rates predicted by linear stability theory~\cite{youngs1984numerical} during the early stages of instability development. The dimensionless growth rates, extracted 
from interface amplitude evolution, were consistent with theoretical predictions and benchmark numerical studies~\cite{youngs1984numerical}. In the nonlinear regime, the simulation captured the emergence of roll-up and finger-like structures, in agreement with high-resolution studies of RTI dynamics, thereby demonstrating the method’s robustness in handling topological changes and interface reconnection without explicit interface tracking.

The NSCH framework proved to be a versatile and robust approach for modeling immiscible two-phase flows with complex interface dynamics, large deformations, and topological changes. Its diffuse-interface formulation naturally accommodates interfacial curvature effects, breakup, and coalescence, making it well-suited for multiphysics applications where phase separation couples strongly with fluid motion. While this work focused on isothermal conditions, the methodology can be extended to incorporate thermal effects, enabling the investigation of thermally driven phase transformations and boiling phenomena. Such extensions are particularly relevant to microelectronics cooling \cite{lorenzini2019numerical,lorenzini2024numerical}, where localized heating induces bubble nucleation and departure, significantly affecting heat transfer efficiency. The present formulation, with appropriate coupling to an energy equation, offers a direct path toward predictive simulations of such thermofluidic processes.

In summary, this study demonstrates that the phase-field method, implemented within the NSCH system, is capable of delivering both qualitative and quantitative agreement with established theory and experiments for immiscible two-phase flows. The findings not only validate the computational framework but also position it as a powerful tool for future research into multiphysics problems involving interfacial dynamics, thermal effects, and complex geometries relevant to industrial and microscale applications.

\section{Appendix}
\begin{tcolorbox}[colback=gray!10, colframe=gray!50, title=Taylor Series Expansion]

The Taylor series expansion of a scalar function $f$ about a point $x_0$ is given by
\begin{align}
f(x) &= f(x_0) 
+ \frac{\partial f}{\partial x}\Big|_{x=x_0} (x - x_0) \nonumber \\
&\quad + \frac{1}{2!} \frac{\partial^2 f}{\partial x^2}\Big|_{x=x_0} (x - x_0)^2 
+ \cdots
\label{eq:taylor}
\end{align}

For a multivariable function expanded about the point 
$\mathbf{x_0} = (x_{1,0}, x_{2,0}, \ldots, x_{n,0})$, we have
\begin{align}
f(\mathbf{x}) &= f(\mathbf{x_0}) 
+ \sum_i \frac{\partial f}{\partial x_i}\Big|_{\mathbf{x}=\mathbf{x_0}} (x_i - x_{i,0}) \nonumber \\
&\quad + \frac{1}{2!} \sum_{i,j} \frac{\partial^2 f}{\partial x_i \partial x_j}
\Big|_{\mathbf{x}=\mathbf{x_0}} (x_i - x_{i,0})(x_j - x_{j,0}) \nonumber \\
&\quad + \cdots
\end{align}

Here, $\mathbf{x} = (x_1, x_2, \ldots, x_n)$ denotes the vector of variables and 
$\mathbf{x_0} = (x_{1,0}, x_{2,0}, \ldots, x_{n,0})$ is the expansion point.

\end{tcolorbox}

\subsection{\label{sec:free_energy}Expression for Free Energy Functional}
This section outlines the development of a general equation for free energy in nonuniform systems, crucial for modeling complex interfacial phenomena. Such systems exhibit spatial variations in the phase field, and the local free energy density \( f \) depends not only on the local phase field variable \( \phi \) but also on its gradients. To capture this, we decompose \( f \) into two contributions:  
(1) a function of the local phase field variable, and  
(2) functions of its derivatives.  
This approach ensures both compositional and gradient effects are represented.  
To formalize this idea, we expand \( f \) in a Taylor series around the reference state \( f_0(\phi) \), corresponding to a uniform phase field (\(\nabla \phi = 0\)) \cite{cahn1961spinodal}.

\paragraph{Taylor Expansion of Free Energy}
\begin{align}
f(\phi, \nabla \phi, \nabla^2 \phi, \ldots) 
&= f_0(\phi) \nonumber \\[6pt]
&\quad + \sum_i 
   \left( 
     \frac{\partial f}{\partial \left( \frac{\partial \phi}{\partial x_i} \right)} 
   \right)_0 
   \left( \frac{\partial \phi}{\partial x_i} \right) \nonumber \\[6pt]
&\quad + \sum_{ij} 
   \left( 
     \frac{\partial f}{\partial \left( \frac{\partial^2 \phi}{\partial x_i \partial x_j} \right)} 
   \right)_0 
   \left( \frac{\partial^2 \phi}{\partial x_i \partial x_j} \right) \nonumber \\[6pt]
&\quad + \tfrac{1}{2} \sum_{ij} 
   \Bigg[
   \begin{aligned}[t]
     &\left( 
       \frac{\partial^2 f}
       {\partial \left( \frac{\partial \phi}{\partial x_i} \right) 
        \partial \left( \frac{\partial \phi}{\partial x_j} \right)} 
     \right)_0 \\[6pt]
     &\times 
     \left( \frac{\partial \phi}{\partial x_i} \right) 
     \left( \frac{\partial \phi}{\partial x_j} \right)
   \end{aligned}
   \Bigg] \nonumber \\[6pt]
&\quad + \cdots
\label{eq:free_taylor}
\end{align}

\paragraph{Coefficient Definitions}
To simplify notation, we define:
\begin{align}
L_i &= 
\left( \frac{\partial f}{\partial \left( \frac{\partial \phi}{\partial x_i} \right)} \right)_0, 
\label{li} \\
\kappa_{ij}^{(1)} &= 
\left( \frac{\partial f}{\partial \left( \frac{\partial^2 \phi}{\partial x_i \partial x_j} \right)} \right)_0, 
\label{k1} \\
\kappa_{ij}^{(2)} &= 
\left( \frac{\partial^2 f}
{\partial \left( \frac{\partial \phi}{\partial x_i} \right) 
 \partial \left( \frac{\partial \phi}{\partial x_j} \right)} \right)_0. 
\label{k2}
\end{align}
\paragraph{Final Expression}
Using these definitions, the expansion of \( f \) is:
\begin{align}
f(\phi, \nabla \phi, \nabla^2 \phi, \ldots) &= f_0(\phi) \notag \\
&\quad + \sum_i L_i 
  \left( \frac{\partial \phi}{\partial x_i} \right) \notag \\
&\quad + \sum_{ij} \kappa_{ij}^{(1)} 
  \left( \frac{\partial^2 \phi}{\partial x_i \partial x_j} \right) \notag \\
&\quad + \frac{1}{2} \sum_{ij} \kappa_{ij}^{(2)} 
  \left( \frac{\partial \phi}{\partial x_i} \right) 
  \left( \frac{\partial \phi}{\partial x_j} \right) + \cdots
\label{eq:free_energy_expansion}
\end{align}

\paragraph{Symmetry Constraints}
In general, \(\kappa_{ij}^{(1)}\) and \(\kappa_{ij}^{(2)}\) are tensors reflecting the symmetry of the fluid mixture, while \( L_i \) represents gradient-dependent contributions.  
For an isotropic and homogeneous binary mixture, the system must remain invariant under spatial reflections and rotations.  

Consider a reflection about the \(x_i\)-axis:
\[
x_i \rightarrow -x_i, 
\qquad 
\frac{\partial \phi}{\partial x_i} \rightarrow -\frac{\partial \phi}{\partial x_i}.
\]
The first-order term transforms as:
\begin{align}
\sum_i L_i \left( \frac{\partial \phi}{\partial x_i} \right) 
&\rightarrow 
\sum_i L_i \left( -\frac{\partial \phi}{\partial x_i} \right) \nonumber \\
&= - \sum_i L_i \left( \frac{\partial \phi}{\partial x_i} \right).
\end{align}
For invariance, this requires
\begin{equation}
\boxed{L_i = 0}.
\end{equation}
Thus, gradient-linear contributions vanish by symmetry, and the leading correction to \( f \) arises from quadratic gradient terms.

\subsubsection{Derivation of \(\kappa_{ij}^{(1)}\)}

\paragraph{Symmetry Conditions on \(\kappa_{ij}^{(1)}\)}  

The coefficients \(\kappa_{ij}^{(1)}\) and \(\kappa_{ij}^{(2)}\) must respect the symmetry and isotropy of a binary fluid mixture.  
These requirements ensure that the free energy functional remains invariant under spatial transformations such as rotations and reflections.  

The term involving \(\kappa_{ij}^{(1)}\) in the Taylor series expansion is given by:
\begin{equation}
   \sum_{ij} \kappa_{ij}^{(1)} 
   \left( \frac{\partial^2 \phi}{\partial x_i \partial x_j} \right).
\end{equation}

For an isotropic and homogeneous binary fluid mixture, invariance under rotations and reflections implies that:  

\begin{align}
   \kappa_{ij}^{(1)} &= \kappa_1, \quad && i = j, \\[6pt]
   \kappa_{ij}^{(1)} &= 0, \quad && i \ne j.
\end{align}

\paragraph{Definition of \(\kappa_1\)}  
The isotropic coefficient \(\kappa_1\) is defined as the derivative of the free energy density with respect to the Laplacian of \(\phi\), evaluated at the reference state:
\begin{equation}
   \kappa_1 
   = \left[ \frac{\partial f}{\partial (\nabla^2 \phi)} \right]_0.
\end{equation}

Therefore, the conditions for \(\kappa_{ij}^{(1)}\) can be summarized compactly as:  

\begin{equation}
\boxed{
\kappa_{ij}^{(1)} =
\begin{cases}
   \displaystyle \left[ \frac{\partial f}{\partial (\nabla^2 \phi)} \right]_0, & i = j, \\[12pt]
   0, & i \ne j.
\end{cases}
}
\end{equation}

\subsubsection{Derivation of \(\kappa_{ij}^{(2)}\)} 

\paragraph{The Term Involving \(\kappa_{ij}^{(2)}\)}  

The term involving \(\kappa_{ij}^{(2)}\) in the Taylor series expansion is:  

\begin{equation}
   \frac{1}{2} \sum_{ij} \kappa_{ij}^{(2)} 
   \left[ \left( \frac{\partial \phi}{\partial x_i} \right) 
   \left( \frac{\partial \phi}{\partial x_j} \right) \right].
\end{equation}

Similar to \(\kappa_{ij}^{(1)}\), the coefficients \(\kappa_{ij}^{(2)}\) must be identical for all diagonal elements and vanish for off-diagonal elements in order to maintain invariance under rotation and reflection.  

\paragraph{Definition of \(\kappa_2\)}  
The coefficient \(\kappa_2\) is defined as the second derivative of the free energy density with respect to \((\partial \phi)^2\), evaluated at the reference state:  

\begin{equation}
   \kappa_2 = \left[ \frac{\partial^2 f}{\partial (\partial \phi)^2} \right]_0.
\end{equation}

Thus, the conditions for \(\kappa_{ij}^{(2)}\) are:  

\begin{equation}
\boxed{
\begin{aligned}
\kappa_{ij}^{(2)} &= \kappa_2 = \left[ \frac{\partial^2 f}{\partial (\partial \phi)^2} \right]_0, 
&& \text{for } i = j, \\[6pt]
\kappa_{ij}^{(2)} &= 0, 
&& \text{for } i \ne j.
\end{aligned}
}
\end{equation}

\paragraph{Free Energy Expression}  
Integrating over a volume \(V\) of the solution, we obtain the total free energy:  

\begin{align}
F &= \int_V f \, dV, \\
  &= \int_V \left[ f_0(\phi) + \kappa_1 \nabla^2 \phi 
     + \kappa_2 (\nabla \phi)^2 + \cdots \right] dV. 
     \label{eq:free_energy_expansion1}
\end{align}

Hence, for a binary fluid mixture, the free energy density reduces to:  

\begin{equation}
\boxed{
f(\phi, \nabla \phi, \nabla^2 \phi, \ldots) 
= f_0(\phi) + \kappa_1 \nabla^2 \phi + \kappa_2 (\nabla \phi)^2 + \cdots
}
\end{equation}

\paragraph{Simplifying the Integral Term}  
Consider the contribution:  
\begin{equation}
\int_V (\kappa_1 \nabla^2 \phi) \, dV. \label{eq5}
\end{equation}

Since \(\kappa_1\) is a function of \(\phi\), we apply the product rule:  
\begin{equation}
\kappa_1 \nabla^2 \phi 
= \nabla \cdot (\kappa_1 \nabla \phi) - \nabla \kappa_1 \cdot \nabla \phi.
\end{equation}

Substituting into Eq.~\eqref{eq5}, we obtain:  
\begin{equation}
\int_V (\kappa_1 \nabla^2 \phi) \, dV 
= \int_V \nabla \cdot (\kappa_1 \nabla \phi) \, dV 
- \int_V (\nabla \kappa_1 \cdot \nabla \phi) \, dV.
\end{equation}

\begin{tcolorbox}[colback=gray!10, colframe=gray!50, title=Divergence Theorem]
The divergence theorem (Gauss’s theorem) relates the flux of a vector field through a closed surface to the divergence of the field within the enclosed volume:
\[
\int_V (\nabla \cdot \mathbf{A}) \, dV 
= \int_S (\mathbf{A} \cdot \mathbf{n}) \, dS,
\]
where \(V\) is the volume, \(S\) is the enclosing surface, \(\mathbf{A}\) is a vector field, and \(\mathbf{n}\) is the outward normal unit vector on \(S\).
\end{tcolorbox}

Applying this to the first term yields:  
\begin{equation}
\int_V \nabla \cdot (\kappa_1 \nabla \phi) \, dV 
= \int_S (\kappa_1 \nabla \phi \cdot \mathbf{n}) \, dS.
\end{equation}

Thus,  
\begin{equation}
\int_V (\kappa_1 \nabla^2 \phi) \, dV 
= \int_S (\kappa_1 \nabla \phi \cdot \mathbf{n}) \, dS 
- \int_V (\nabla \kappa_1 \cdot \nabla \phi) \, dV.
\end{equation}

\paragraph{Simplifying the Second Term}  
Since \(\kappa_1 = \kappa_1(\phi)\):  
\begin{equation}
\nabla \kappa_1 = \frac{d\kappa_1}{d\phi} \nabla \phi,
\end{equation}
which gives:  
\begin{equation}
\nabla \kappa_1 \cdot \nabla \phi 
= \left( \frac{d\kappa_1}{d\phi} \right) (\nabla \phi)^2.
\end{equation}

Substituting back, we obtain:  
\begin{equation}
\int_V (\kappa_1 \nabla^2 \phi) \, dV 
= \int_S (\kappa_1 \nabla \phi \cdot \mathbf{n}) \, dS 
- \int_V \left( \frac{d\kappa_1}{d\phi} \right) (\nabla \phi)^2 \, dV.
\end{equation}

Finally, neglecting boundary effects by choosing \( \nabla \phi \cdot \mathbf{n} = 0 \) at the boundary, the surface integral vanishes, yielding:  
\begin{equation}
\int_V (\kappa_1 \nabla^2 \phi) \, dV 
= - \int_V \left( \frac{d\kappa_1}{d\phi} \right) (\nabla \phi)^2 \, dV.
\end{equation}

\paragraph{Final Expression for the Free Energy Functional}  
Substituting this back into Eq.~\eqref{eq:free_energy_expansion1}, we obtain:  

\begin{equation}
\boxed{
F[\phi] = \int_V \left[ f_0(\phi) + \kappa (\nabla \phi)^2 + \cdots \right] dV
} \label{cahn_expression}
\end{equation}

where
\begin{align}
\kappa &= - \frac{d \kappa_1}{d \phi} + \kappa_2, \\
       &= - \left[ \frac{\partial^2 f}{\partial \phi \, \partial (\nabla^2 \phi)} \right]_0
          + \left[ \frac{\partial^2 f}{\partial (\partial \phi)^2} \right]_0.
\end{align}

Equation~\eqref{cahn_expression} is pivotal: it shows that, to first approximation, the free energy of a small volume of a nonuniform solution can be expressed as the sum of two contributions—the homogeneous free energy density \(f_0\) and a ``gradient energy'' term that depends on the local order parameter field \(\phi\).

\subsection{\label{sec:CH-Adv Appendix} Advective Cahn–Hilliard Formulation}

In this section, we derive the Cahn–Hilliard equation in the presence of fluid motion by incorporating an advective term. This extended form, denoted as Eq.~\ref{CH:adv} in Section~\ref{sec:CHEq}, is essential when modeling mixtures of incompressible and immiscible fluids where transport occurs not only by diffusion but also by bulk advection. To illustrate, we consider two fluids, A and B, with distinct densities and viscosities. Their local composition within a control volume can be represented through a volume fraction approach, akin to the Volume of Fluid (VOF) method. Specifically, we use the volume fraction of fluid A, denoted by \( C \) (\(0 \leq C \leq 1\)), as the primary descriptor of the mixture.

\paragraph*{Density definitions.}  
Given \(C\), the local densities of fluids A and B in a volume element are:
\begin{equation}
\tilde{\rho}_A = C \rho_A, 
\quad 
\tilde{\rho}_B = (1 - C)\rho_B,
\end{equation}
where \(\rho_A\) and \(\rho_B\) are the bulk densities. The average density is thus:
\[
\rho = C \rho_A + (1 - C)\rho_B.
\]

\paragraph*{Mass conservation.}  
To establish the continuity equation, consider the mass of species A within a control volume \(V\):
\[
M_A = \int_V \tilde{\rho}_A \, dV.
\]
Its time rate of change is
\[
\frac{d}{dt}\int_V \tilde{\rho}_A \, dV.
\]
Applying the divergence theorem, the rate of change must balance the net flux across the surface \(S\) bounding \(V\):
\[
\frac{d}{dt}\int_V \tilde{\rho}_A \, dV 
= -\int_S \mathbf{n}_A \cdot \mathbf{n}\, dS
= -\int_V \nabla \cdot \mathbf{n}_A \, dV,
\]
where \(\mathbf{n}_A\) is the flux of species A and \(\mathbf{n}\) the outward normal. Since \(V\) is arbitrary:
\begin{equation}
\boxed{
\frac{\partial \tilde{\rho}_A}{\partial t} + \nabla \cdot \mathbf{n}_A = 0
}\label{cont}
\end{equation}

\paragraph*{Flux decomposition.}  
The total flux of species A comprises advective and diffusive contributions:
\[
\mathbf{n}_A = \underbrace{\tilde{\rho}_A \mathbf{u}}_{\text{advection}} 
- \underbrace{\rho_A \mathbf{j}_A}_{\text{diffusion}},
\]
where \(\mathbf{u}\) is the fluid velocity and \(\mathbf{j}_A\) is the diffusive flux. Neglecting diffusion (\(\mathbf{j}_A=0\)) reduces \(\mathbf{n}_A\) to its advective form.

\paragraph*{Continuity equation for volume fraction.}  
Substituting $\tilde{\rho}_A = C \rho_A$ into Eq.~\eqref{cont} gives
\begin{align}
\frac{\partial (C \rho_A)}{\partial t} 
&+ \nabla \cdot (C \rho_A \mathbf{u})
- \nabla \cdot (\rho_A \mathbf{j}_A) = 0.
\end{align}

Normalizing by $\rho_A$ yields
\begin{align}
\boxed{
\frac{\partial C}{\partial t}
+ \nabla \cdot (C \mathbf{u})
- \nabla \cdot \mathbf{j}_A = 0
}
\label{eq:component_A}
\end{align}
\noindent\textit{for component A.}

Similarly, for fluid B we obtain
\begin{align}
\boxed{
\frac{\partial (1 - C)}{\partial t}
+ \nabla \cdot \big[(1 - C)\,\mathbf{u}\big]
- \nabla \cdot \mathbf{j}_B = 0
}
\label{eq:component_B}
\end{align}
\noindent\textit{for component B.}

\paragraph*{Divergence-free velocity condition.}  
Adding Eqs.~\eqref{eq:component_A} and \eqref{eq:component_B} gives
\[
\nabla \cdot \mathbf{u} = \nabla \cdot (\mathbf{j}_A + \mathbf{j}_B).
\]
Since interfacial diffusion ensures
\begin{equation}
\mathbf{j}_B = -\mathbf{j}_A,
\label{flux}
\end{equation}
we obtain
\[
\nabla \cdot \mathbf{u} = 0,
\]
which enforces incompressibility of the velocity field.

\paragraph*{Final advective Cahn–Hilliard equation.}  
The governing evolution equation for the volume fraction \(C\) becomes
\begin{equation}
\boxed{
\frac{\partial C}{\partial t} + \mathbf{u} \cdot \nabla C 
- \nabla \cdot \mathcal{J} = 0
}\label{ch_ad}
\end{equation}
where \(\mathcal{J} = \mathbf{j}_A = -\mathbf{j}_B\) is the diffusive flux.  

By replacing the volume fraction \(C\) with the order parameter \(\phi\), which ranges from \(-1\) (pure B) to \(+1\) (pure A), Eq.~\eqref{ch_ad} transforms into the convective Cahn–Hilliard equation:
\begin{equation}
\phi_t + \mathbf{u} \cdot \nabla \phi = \nabla \cdot \mathcal{J}.
\label{CH:adv}
\end{equation}
}
This formulation is general and remains valid even for fluids with different densities and viscosities, since these contrasts do not alter the fundamental structure of the Cahn–Hilliard equation. The order parameter thus serves as a natural descriptor of the fluid mixture, linking advective transport with interfacial diffusion in a unified framework.

\subsection{Derivation of Linear Interpolation Formulae}
\label{app:interp}

\subsubsection{Viscosity interpolation}
For viscosity, set $(x_0,y_0) = (-1, \eta_1)$ and $(x_1,y_1) = (1, \eta_2)$:
\begin{align}
\eta(\phi) &= \eta_1 + \frac{\phi - (-1)}{1 - (-1)} (\eta_2 - \eta_1) \notag \\
&= \frac{\phi + 1}{2} \, \eta_2 + \frac{1 - \phi}{2} \, \eta_1 \notag \\
&= \frac{\phi + 1}{2} (\eta_2 - \eta_1) + \eta_1.
\end{align}
Replacing $\phi$ with $\hat{\phi}$ gives Eq.~\eqref{eq:eta_interp}

\begin{tcolorbox}[colback=gray!10, colframe=gray!50, title=Linear Interpolation]
Linear interpolation estimates a value between two known points $(x_0, y_0)$ and $(x_1, y_1)$ using:
\[
y = y_0 + \frac{x - x_0}{x_1 - x_0} \, (y_1 - y_0).
\]
\end{tcolorbox}
\subsubsection{Density interpolation}
Similarly, for density $(\rho_1,\rho_2)$:
\begin{align}
\rho(\phi) &= \rho_1 + \frac{\phi - (-1)}{1 - (-1)} (\rho_2 - \rho_1) \notag \\
&= \frac{\phi + 1}{2} \, \rho_2 + \frac{1 - \phi}{2} \, \rho_1 \notag \\
&= \frac{\phi + 1}{2} (\rho_2 - \rho_1) + \rho_1,
\end{align}
which matches Eq.~\eqref{eq:rho_interp} after truncation.

\section*{Acknowledgements}
This research was supported by the U.S. Department of Energy (DOE), Office of Science, Office of Basic Energy Sciences, through the Data, Artificial Intelligence, and Machine Learning at DOE Scientific User Facilities program under Award No. 34532 (Digital Twins). Portions of this work were carried out at the Center for Nanoscale Materials and the Advanced Photon Source, both DOE Office of Science User Facilities supported by the Office of Basic Energy Sciences, under Contracts DE-AC02-06CH11357 and DE-AC05-00OR22725, respectively. Computational resources were provided by the National Energy Research Scientific Computing Center (NERSC), a DOE Office of Science User Facility supported under Contract DE-AC02-05CH11231, as well as by the Laboratory Computing Resource Center (LCRC) at Argonne National Laboratory.

\bibliographystyle{apsrev4-2} 
\bibliography{references}     

\end{document}